\documentclass[final,3p,times,twocolumn]{elsarticle}

\usepackage{graphicx}
\usepackage{txfonts}
\usepackage{caption}
\usepackage{mathabx}
\usepackage{lineno}

\usepackage{caption}
\usepackage{subcaption}
\usepackage{mathabx}
\usepackage[hidelinks]{hyperref}
\usepackage{array, blkarray, makecell}
\usepackage{tabularx}
\usepackage{pdflscape}
\usepackage{afterpage}
\usepackage{wasysym}
\usepackage{color}

\biboptions{authoryear}

\begin{document}

\begin{frontmatter}



\title{Mineralogy, structure and habitability of carbon-enriched rocky exoplanets: A laboratory approach}


\author[{UvA,VU}]{Kaustubh Hakim*}
\author[ETH]{Rob Spaargaren}
\author[Rice]{Damanveer S. Grewal}
\author[WWU]{Arno Rohrbach}
\author[WWU]{Jasper Berndt}
\author[UvA]{Carsten Dominik}
\author[VU]{Wim van Westrenen}

\address[UvA]{Anton Pannekoek Institute for Astronomy, University of Amsterdam, Science Park 904,
1098 XH Amsterdam, The Netherlands}
\address[VU]{Department of Earth Sciences, Vrije Universiteit, De Boelelaan 1085, 
1081 HV Amsterdam, The Netherlands}
\address[ETH]{Department of Earth Sciences, ETH Z{\"u}rich, Sonneggstrasse 5, 8092 Z{\"u}rich, Switzerland}
\address[Rice]{Department of Earth, Environmental and Planetary Sciences, Rice University, MS 126, 6100 Main Street, Houston, TX 77005, USA}
\address[WWU]{Institut f{\"u}r Mineralogie, Westf{\"a}lische Wilhelms$–$Universit{\"a}t M{\"u}nster, Corrensstrasse 24, 48149 M{\"u}nster, Germany}

%Accepted, 9 May 2018. Received, 2 August 2017.

\journal{*Corresponding author email: hakim.kaustubh@gmail.com}

\begin{abstract}

Carbon-enriched rocky exoplanets have been proposed around dwarf stars as well as around binary stars, white dwarfs and pulsars. However, the mineralogical make up of such planets is poorly constrained.  We performed high-pressure high-temperature laboratory experiments ($P$ = 1$-$2~GPa, $T$ = 1523$-$1823~K) on chemical mixtures representative of C-enriched rocky exoplanets based on calculations of protoplanetary disk compositions. These $P-T$ conditions correspond to the deep interiors of Pluto- to Mars-size planets and the upper mantles of larger planets.

Our results show that these exoplanets, when fully-differentiated, comprise a metallic core, a silicate mantle and a graphite layer on top of the silicate mantle. Graphite is the dominant carbon-bearing phase at the conditions of our experiments with no traces of silicon carbide or carbonates. The silicate mineralogy comprises olivine, orthopyroxene, clinopyroxene and spinel, similar to the mineralogy of the mantles of carbon-poor planets such as the Earth, and largely unaffected by the amount of carbon. Metals are either two immiscible iron-rich alloys (S-rich and S-poor) or a single iron-rich alloy in the Fe-C-S system with immiscibility depending on the S/Fe ratio and core pressure. 
   
We show that for our C-enriched compositions the minimum carbon abundance needed for C-saturation is 0.05$-$0.7~wt\% (molar C/O $\sim$ 0.002$-$0.03). Fully differentiated rocky exoplanets with C/O ratios more than needed for C-saturation would contain graphite as an additional layer on top of the silicate mantle. For a thick enough graphite layer, diamonds would form at the bottom of this layer due to high pressures. 

We model the interior structure of Kepler-37b and show that a mere 10~wt\% graphite layer would decrease its derived mass by 7\%, suggesting future space missions that determine both radius and mass of rocky exoplanets with insignificant gaseous envelopes could provide quantitative limits on their carbon content. Future observations of rocky exoplanets with graphite-rich surfaces would show low albedos due to the low reflectance of graphite. The absence of life-bearing elements other than carbon on the surface likely makes them uninhabitable. 

\end{abstract}

\begin{keyword}
Carbon-rich \sep Rocky Exoplanets \sep Mineralogy \sep High-pressure \sep Laboratory Experiments \sep Habitability
\end{keyword}

\end{frontmatter}

%\linenumbers

\section{Introduction}\label{paper2:introduction}

Since the discovery of the first rocky exoplanet, CoRoT-7b \citep{Leger2009}, more than a thousand such planets have been found\footnote{http://exoplanetarchive.ipac.caltech.edu}. The scatter in the mass-radius diagram of rocky exoplanets reveals a great diversity in their bulk composition and interior structure \citep[e.g.,][]{Valencia2006,Seager2007,Wagner2011,Hakim2018a}. Water/ices \citep[e.g., GJ~876d,][]{Valencia2007a}, thick atmospheres \citep[e.g., GJ~1132b,][]{Southworth2017} as well as carbon-bearing minerals including graphite and silicon carbide \citep[e.g., 55~Cancri~e,][]{Madhusudhan2012} have been suggested as dominant phases in these exoplanets, in addition to silicate minerals and iron alloys. Besides dwarf stars, carbon-enriched rocky exoplanets have been proposed around binary stars \citep[e.g.,][]{Whitehouse2018} as well as pulsars and white dwarf stars \citep[e.g.,][]{Kuchner2005}.

Although life as we know it is largely based on carbon, the Earth contains less than 0.01~wt\% carbon \citep[e.g.,][]{Javoy2010}. This is to some extent surprising, since many weakly processed planetary building blocks in solar system contain significant amounts of carbon in the form of organics \citep[e.g., carbonaceous chondrites,][]{Marty2013}, and some larger, more evolved bodies such as the ureilite parent body contain significant amounts of refractory carbon \citep{Nabiei2018}. In order to explain the extremely low abundance of carbon in the Earth, carbon needs to be burned (oxidized and turned into CO/CO$_{2}$) or photolyzed away (broken out of the organic compounds by energetic photons) while the solid material is still present in the form of small grains with large surface-to-mass ratio \citep{Lee2010,Anderson2017}. 

An alternative explanation is that planetesimal-sized parent bodies need to be subjected to igneous processing to degas carbon \citep{Hashizume1998} which is tied to the presence of radioactive elements like $^{26}$Al in the early solar nebula \citep{Hevey2006}. Both these processes do not seem inevitable. Oxidation and photo-processing may be quenched by effects of dust growth and transport in disks \citep{Klarmann2018}. The presence of short-lived radioactive isotopes in significant amounts requires the fast (within a Myr) addition of the ejecta of a nearby supernova explosion or stellar wind material into the collapsing protosolar cloud, followed by rapid formation of planetesimals \citep{Bizzarro2005}. It is therefore likely that the conditions needed to decarbonize solids are absent in many planet-forming systems, and that rocky planets in such systems may contain significant levels of carbon up to 10 mass percent. 

Even larger carbon abundance could be obtained in systems where the carbon-to-oxygen abundance ratio is higher than in the solar system. Modeling of the protoplanetary disk chemistry for planet-hosting stars with molar photospheric C/O {\textgreater} 0.65 \citep{Moriarty2014} and C/O {\textgreater} 0.8 \citep{Bond2010b,CarterBond2012b} (cf. C/O$_{\mathrm{Sun}}\sim0.54$) suggests that carbon acts as a refractory element mainly in the form of graphite and silicon carbide in the inner regions of such disks.  \citet{DelgadoMena2010} and \citet{Petigura2011} reported spectroscopic observations of stars with photospheric C/O ratios greater than unity. However, \citet{Nakajima2016} and \citet{Brewer2016} claim that the stars in solar neighborhood have largely solar-like C/O ratios. Although the debate over photospheric C/O ratios is not settled, the possibility of a substantial fraction of stars with C/O {\textgreater} 0.65 cannot be excluded.

Refractory elements in protoplanetary disks are the major building blocks of rocky planets. \citet{Bond2010b} found that the C/O ratio of the refractory material in inner disks of stars with C/O {\textgreater} 0.8 varies from zero to greater than one hundred as a function of distance from the star. \citet{Moriarty2014} found that, for high C/O stars, the extent of refractory carbon in the planetesimal disk increases using a sequential condensation model instead of a simple equilibrium condensation model. \citet{Thiabaud2015} showed that C/O ratios of rocky planets do not necessarily show a one-to-one correlation with the stellar photospheric C/O ratios. N-body simulations by \citet{Bond2010b} produce rocky exoplanets containing as high as 70~wt\% carbon. The amount and nature of carbon-bearing minerals in carbon-enriched rocky exoplanets may directly impact geodynamical processes, carbon and water cycles and, in turn planetary habitability \citep{Unterborn2014}. 

During the early stages of planet formation, refractory material in protoplanetary disks condenses out from the chemical reactions between gas molecules. Coagulation of refractory material leads to the formation of pebbles, which grow into sub-Ceres-size to Pluto-size planetesimals and later on form planets \citep{Johansen2007,SchaferJohansen2017}. Such planetesimals are large enough to undergo large-scale differentiation at high-pressure-temperature conditions during the process of planet formation. Modeling studies such as \cite{Bond2010b} and \citet{Moriarty2014} derive proportions of chemical compounds condensing out from gas chemistry and perform N-body simulations on planetesimals to track the likely chemical composition of resulting planets. Since the pressures in interiors of planetesimals and planets are several orders of magnitude higher than the disk pressures, high-pressure high-temperature reactions are expected to reprocess their chemical composition and kick off large-scale differentiation processes in their interiors which lead to metal segregation and core formation \citep{Kruijer2013}. Current understanding of the mineralogy of exoplanets is based on extrapolation of the knowledge of rocky bodies in our solar system and lacks experimental evidence. There is a need to investigate the mineralogy and phase relations of carbon-rich planetesimals and exoplanets, which have no solar system analogs, in multi-component systems, and high-pressure high-temperature experiments make it possible \citep[e.g.,][]{Valencia2009,Nisr2017}.

C-enriched rocky exoplanets are speculated to contain large amounts of C-bearing minerals including silicon carbide and graphite \citep[e.g.,][]{Bond2010b,Madhusudhan2012}. Over the past decades, several laboratory studies have investigated the mineralogy of rocky planets in C-poor Earth-like conditions, but only a few studies are applicable to conditions relevant to C-enriched exoplanetary interiors. \citet{Corgne2008} used a CI-chondrite-like composition to probe early planetesimal differentiation in carbon- and sulfur-enhanced environments and observed liquid metal immiscibility leading to the formation of C-rich and S-rich metals. The extent of liquid metal immiscibility has been explored in the simple Fe-C-S \citep[e.g.,][]{Dasgupta2009}, Fe-S-O \citep[e.g.,][]{Tsuno2007} and Fe-S-Si \citep[e.g.,][]{Morard2010} systems. The solubility of carbon in iron alloys \citep[e.g.,][]{Lord2009,Tsuno2015} and silicate melts  \citep[e.g.,][]{Duncan2017}, the partitioning of carbon between silicate melt and iron alloys \citep[e.g.,][]{Chi2014,Li2015,Li2016} and the stability of reduced versus oxidized carbon in the Earth's mantle \citep[e.g.,][]{Rohrbach2011} have also been investigated. Phase relations have been studied in the carbon-saturated Fe-Mg-Si-C-O (FMS+CO) system with bulk compositions depleted in oxygen \citep{Takahashi2013}. The study by \citet{Takahashi2013} covers a range of oxygen fugacities resembling highly reducing conditions, however they did not consider the presence of S which can be a major component in the Fe cores of rocky bodies \citep[e.g.,][]{Stewart2007,Rai2013,Steenstra2016}. Moreover, they lack a discussion about the diversification of silicate minerals due to the absence of Al and Ca in their experiments. Finally, to our knowledge no experimental studies have used C-enriched starting compositions calculated by modeling planet formation chemistry around stars other than our Sun, which is key for future exoplanetary exploration. 

Here we probe the mineralogy and structure of small C-enriched rocky exoplanets by performing high-pressure high-temperature laboratory experiments on chemical mixtures in the Fe-Ca-Mg-Al-Si-C-S-O (FCMAS+CSO) system resembling the bulk compositions of C-enriched planetesimals from the models of \citet{Moriarty2014}. In Sect. \ref{paper2:method}, we give our experimental and analytical methods. Phase relations and compositions of our experimental run products are given in Sect. \ref{paper2:observations}. The mineralogy and structure of C-enriched rocky exoplanets and their dependence on several factors are discussed in Sect. \ref{paper2:interior}. To illustrate the application of our findings, we discuss the implications of assuming a C-enriched interior on the derived mass, future observations and habitability of Kepler-37b, the smallest known exoplanet till date in Sect. \ref{paper2:kepler37b}. Finally, we summarize our findings and conclusions in Sect. \ref{paper2:conclusion}.

\section{Methods}\label{paper2:method}

\subsection{Choice of bulk compositions}\label{paper2:methodComposition}

%\afterpage{
%\clearpage
%}

\begin{table*}[!ht]
\caption{\label{tab:StartingMaterial} Planetesimal bulk compositions and starting materials } 
\footnotesize
\begin{center}
\begin{tabular}{l|rrr||l|rrr} \hline \hline \rule[0mm]{0mm}{0mm}
Element & SC & EC & TC & Material & SC & EC & TC \\[1mm]																	
\hline \\[-1mm]
Si (mol\%)     & 11.4 &  9.0 & 15.3 & $\mathrm{SiO_{2}}$ (wt\%)    & 30.1 & 19.0 & 36.7 \\
Mg (mol\%)     & 11.4 &  9.3 & 15.3 & MgO (wt\%)                   & 20.2 & 19.0 & 24.6 \\
O (mol\%)      & 45.8 & 30.0 & 51.2 & FeO$^\dagger$ (wt\%)         & 27.3 & 23.1 &  8.1 \\
Fe (mol\%)     & 11.4 &  7.6 & 12.8 & Fe (wt\%)                    &  2.2 &  0.0 & 14.2 \\
S (mol\%)      &  1.9 &  1.3 &  3.6 & FeS (wt\%)                   &  7.0 &  8.8 & 12.6 \\
Al (mol\%)     &  1.4 &  0.9 &  1.0 & $\mathrm{Al_{2}O_{3}}$ (wt\%)&  3.1 &  2.4 &  2.1 \\
Ca (mol\%)     &  0.7 &  0.5 &  0.8 & CaO$^\dagger$ (wt\%)         &  1.7 &  1.5 &  1.7 \\
C (mol\%)      & 16.0 & 41.4 &  $-$ & C (wt\%)                     &  8.4 & 23.5 &  $-$ \\
C/O (mol/mol)  & 0.35 & 1.38 &  $-$ & SiC (wt\%)                   &  0.0 &  5.6 &  $-$ \\[1mm]
\hline
\end{tabular}
\end{center}\caption*{$^\dagger$ CaO and FeO are obtained from $\mathrm{CaCO_{3}}$ and $\mathrm{Fe_{2}O_{3}}$ after decarbonation and reduction.}
\end{table*}

\begin{table*}[!ht]
\caption{\label{tab:ExperimentConditions} Experimental conditions and run product phases } 
\footnotesize
\begin{center}
\begin{tabular}{l|ccrll} \hline \hline \rule[0mm]{0mm}{0mm}
Run & P & T               & t   & $\log f_{\mathrm{O_{2}}}$ & Run product phases (proportions in wt\%) \\	
& (GPa) & (K) & (h) &  ($\Delta$IW)             & {(Graphite is an additional phase in all runs)}  \\[2mm]															
\hline \\[-2mm]
\textbf{SC} \\
1B1t  & 1 & 1823 & 3.5 & $-$0.5          & Olv (25\%) + SiL (61\%) + SrFeL2 (13\%) \\
1B1p  & 2 & 1823 & 3.5 & $-$0.7          & Olv (27\%) + SiL (56\%) + SrFeL2 (18\%) \\
1B1f  & 1 & 1723 & 4   & $-$0.6          & Olv (38\%) + SiL (46\%) + SrFeL2 (16\%) \\
1B1j  & 2 & 1723 & 4   & $-$0.4          & Olv (44\%) + SiL (46\%) + SrFeL2 (10\%) \\
1B1q  & 1 & 1623 & 20  & $-$0.5          & Olv (68\%) + SiL (21\%) + Spi (0.2\%) + SrFeL2 (10.8\%) \\
1B1w  & 2 & 1623 & 29  & $-$0.3          & Olv (47\%) + SiL (41\%) + SrFeL2 (11\%) \\
1B1r  & 1 & 1545 & 20  & $-$\textit{0.5} & Olv (74\%) + Spi (11\%) + CPx (4\%) + SrFeL2 (11\%) \\
\hline
\textbf{EC} \\
2C1a  & 1 & 1823 & 4   & $-$1.2          & Olv (41\%) + SiL (36\%) + SpFeL (9\%) + SrFeL  (14\%)\\
2C1d  & 2 & 1823 & 4   & $-$1.0          & Olv (25\%) + SiL (15\%) + OPx (36\%) + SpFeL (9\%) + SrFeL  (15\%)\\
2C1e  & 1 & 1723 & 4   & $-$1.2          & Olv (30\%) + SiL (13\%) + OPx (31\%) + SpFeL (9\%) + SrFeL  (17\%)\\
2C1c  & 2 & 1723 & 6   & $-$\textit{0.9} & Olv (36\%) + SiL$^\dagger$ + CPx (41\%) + SpFeL (9\%) + SrFeL  (14\%)\\
\hline
\textbf{TC} \\
1A2y  & 1 & 1823 & 3.5 & $-$1.1          & Olv (38\%) + SiL (39\%) + SpFeL (14\%) + SrFeL (8\%) \\
1A2zc & 2 & 1823 & 6.5 & $-$1.0          & Olv (26\%) + SiL (21\%) + OPx (28\%) + SpFeL (15\%) + SrFeL (10\%) \\
1A2zd & 1 & 1723 & 8   & $-$1.1          & Olv (35\%) + SiL (41\%) + SpFeL (14\%) + SrFeL (9\%)\\
1A2a  & 2 & 1723 & 4   & $-$1.0          & Olv (22\%) + SiL$^\dagger$ + OPx (52\%) + SpFeL (15\%) + SrFeL (11\%)\\
1A2c  & 1 & 1623 & 4   & $-$\textit{1.2} & Olv (18\%) + SiL$^\dagger$ + OPx (56\%) + SpFeL (15\%) + SrFeL (12\%)\\
1A2za & 2 & 1623 & 29  & $-$\textit{1.2} & Olv (28\%) + OPx (46\%) + SpFeL (15\%) + SrFeL (11\%)\\
1A2s  & 1 & 1523 & 100 & $-$\textit{1.0} & Olv (49\%) + OPx (20\%) + CPx (7\%) + FeS (24\%)\\[2mm]
\hline
\end{tabular}
\end{center}\caption*{ \textit{Note:} Oxygen fugacity is calculated assuming a non-ideal solution behavior of S-rich Fe alloy and silicate melt (see \ref{paper2:appendixOxygenFugacity} for details). Oxygen fugacities in italics are calculated using olivine instead of silicate melt. \\
$^\dagger$ Silicate melt was present in small quantities which could not be measured using EPMA. \\
\textit{Abbreviations:} Olv = Olivine, SiL = Silicate melt, OPx = Orthopyroxene, CPx = Clinopyroxene, Spi = Spinel, SpFeL = S-poor Fe melt, SrFeL = S-rich Fe melt, FeS = Fe-S solid (single alloy), SrFeL2 = S-rich Fe melt (single alloy).
}
\end{table*}

To prepare starting materials for our experiments, we used relative elemental abundances of C-enriched planetesimals at 1~AU and 0.15~Myr after disk formation in the HD19994 planetary system calculated by \citet{Moriarty2014} for their equilibrium chemistry (EC) and sequential condensation chemistry (SC) cases. Two end-member compositions (SC and EC) were prepared using elemental proportions given in Table~\ref{tab:StartingMaterial}. The C/O ratios of the SC and EC compositions are 0.35 and 1.38 respectively, about two-three orders of magnitude higher than that of the Earth, and give an appropriate range of carbon-enriched compositions based on the calculations of \citet{Moriarty2014}. Since our experiments were performed in carbon-saturated conditions by enclosing samples in graphite capsules (see below), there is no upper limit on the amount of carbon in the resulting experiments, and hence these C/O ratios merely signify lower limits. We also chose a third bulk composition (hereafter, TC) resembling solar system terrestrial planetesimals at 1~AU and 0.15~Myr after disk formation from the equilibrium chemistry model of \citet{Moriarty2014}. The TC composition is also saturated with carbon.

\subsection{Starting materials}\label{paper2:methodStartingMaterial}

Starting materials were mixed in proportions shown in Table \ref{tab:StartingMaterial}. In the first step, $\mathrm{SiO_{2}}$ (99.9\% $\mathrm{SiO_{2}}$ powder from Alfa-Aesar), MgO (99.95\% MgO powder from Alfa-Aesar), $\mathrm{Al_{2}O_{3}}$ (99.95\% min alpha $\mathrm{Al_{2}O_{3}}$ powder from Alfa-Aesar), $\mathrm{CaCO_{3}}$ (99.95-100.05\% ACS chelometric standard $\mathrm{CaCO_{3}}$ powder from Alfa-Aesar) and $\mathrm{Fe_{2}O_{3}}$ (99.9\% $\mathrm{Fe_{2}O_{3}}$ powder from Alfa-Aesar) were homogenized in an agate mortar under ethanol. The oxide-carbonate mixture was decarbonated and reduced in a box furnace by first gradually increasing the temperature from 873~K to 1273~K in six hours. The decarbonated mixture, placed in a platinum crucible, was then heated to 1823~K in a box furnace for 30 minutes and then quenched to room temperature by immersing the bottom of the platinum crucible in water, leading to the formation of glassy material. It was then ground to a homogeneous powder using an agate mortar under ethanol. Fe (99.95\% Fe powder, spherical, \textless 10 micron from Alfa-Aesar), FeS (99.9\% FeS powder from Alfa-Aesar), C (99.9995\% Ultra F purity graphite from Alfa-Aesar) and SiC ($\ge$97.5\% SiC powder from Sigma-Aldrich) were added to the powder. The final mixture was again homogenized by grinding in an agate mortar and stored in an oven at 383~K until use.

\subsection{High pressure-temperature experiments}\label{paper2:methodExperiment}

Experiments summarized in Table \ref{tab:ExperimentConditions} were conducted in an end-loaded piston cylinder apparatus at Vrije Universiteit Amsterdam in a 12.7~mm (half-inch) diameter cylindrical sample assembly. Details on sample assembly preparation are given in \ref{paper2:appendixSampleAssembly} and Fig.~\ref{fig:SampleAssembly}. Pressure and temperature conditions of 1$-$2~GPa and 1523$-$1823~K were chosen to represent the interior conditions of Pluto-mass planetesimals and planets. To reduce the porosity of graphite capsules, the sample assembly was sintered at 1073~K and 1~GPa for 1~h before further heating and pressurization. During heating to the run temperature, the pressure was increased continuously with the hot-piston-in technique \citep{McDade2002}. The temperature was increased at a rate of 100~K/min. The experiments were run for the duration of 3.5$-$100~h (Table \ref{tab:ExperimentConditions}). All experiments were quenched to \textless 450~K within $\sim$15~s by switching off the electric power to the heater.

\subsection{Analytical procedure}\label{paper2:methodAnalysis}

The recovered samples were mounted in one-inch-diameter mounts using petropoxy resin, cut longitudinally, polished with grit-paper and fine-polished down to a 1/4~$\mathrm{\mu}$m finish. The polished samples were carbon-coated to ensure electrical conductivity of the surface during electron probe micro-analysis. Major element contents of the experimental charges were determined using wavelength dispersive spectroscopy (WDS) on the 5-spectrometer JEOL JXA-8530F Hyperprobe Electron Probe Micro-Analyzer (EPMA) at the Netherlands National Geological Facility, Utrecht University. We used a series of silicate, oxide and metal standards and conditions of 15~nA beam current and 15 kV accelerating voltage. Analyses were made with a defocused beam to obtain the compositions of metal (2$-$10 $\mathrm{\mu}$m diameter) and silicate (5$-$20~$\mathrm{\mu}$m diameter) phases. Standards for the quantitative analysis of Mg, Fe, Si, Al and Ca in silicate minerals were forsterite, hematite, forsterite, corundum and diopside, respectively and the standard for Fe in iron alloys was Fe-metal. Counting times were 30~s for Fe (hematite and Fe-metal), Si, Mg and Al, and 20~s for Ca and S. Quantitative analysis of Pt, with the help of a Pt-metal standard, was also performed to assess contamination from the Pt capsule. To measure light element abundances in iron alloys, the carbon coating was removed and the samples and standards (natural troilite for S, pure Si metal for Si, magnetite for O and experimentally synthesized Fe$_{3}$C for C) were Al-coated together for each run to keep the X-ray absorption uniform. These analyses were performed using a JEOL JXA 8530F Hyperprobe at Rice University, Houston following the analytical protocol of \citet{Dasgupta2008}. Detection limits (3$\sigma$) of all elements are less than 0.03~wt\% except for Pt (0.07~wt\%). Data reduction was performed using the $\phi$(rZ) correction \citep{Armstrong1995}. Instrument calibrations were deemed successful when the composition of secondary standards was reproduced within the error margins defined by the counting statistics.

\section{Experimental observations}\label{paper2:observations}

\subsection{Phase assemblages and texture}\label{paper2:phaseAssemblages}

\begin{figure*}[!ht]
  \centering
  \medskip
  \includegraphics[width=.8\textwidth]{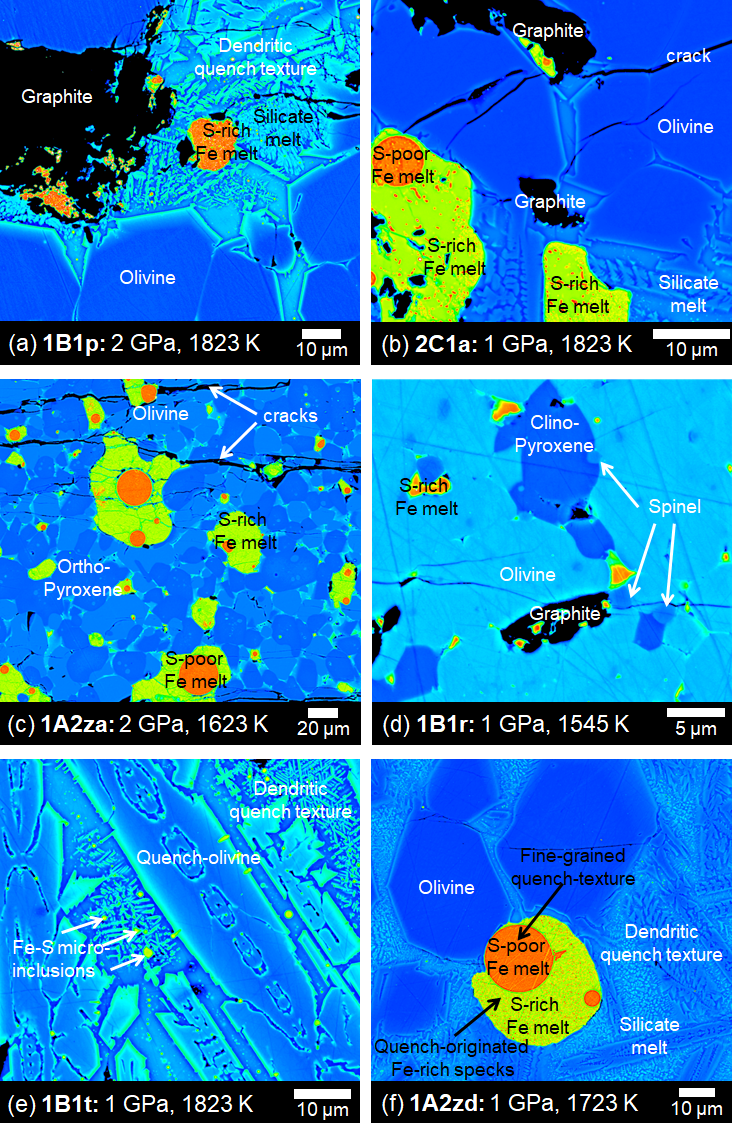}
  \caption[Microprobe Images]{ False-color backscattered electron images of six representative run products illustrating different phases and textural types. Phases can be broadly categorized into graphite, iron alloys and silicate phases. Silicate melts show a typical dendritic quench texture. Iron alloys show a fine-grained quench texture. }
  \label{fig:MicroprobeImages}
\end{figure*}

\begin{figure*}[!ht]
  \centering
  \medskip
  \includegraphics[width=.83\textwidth]{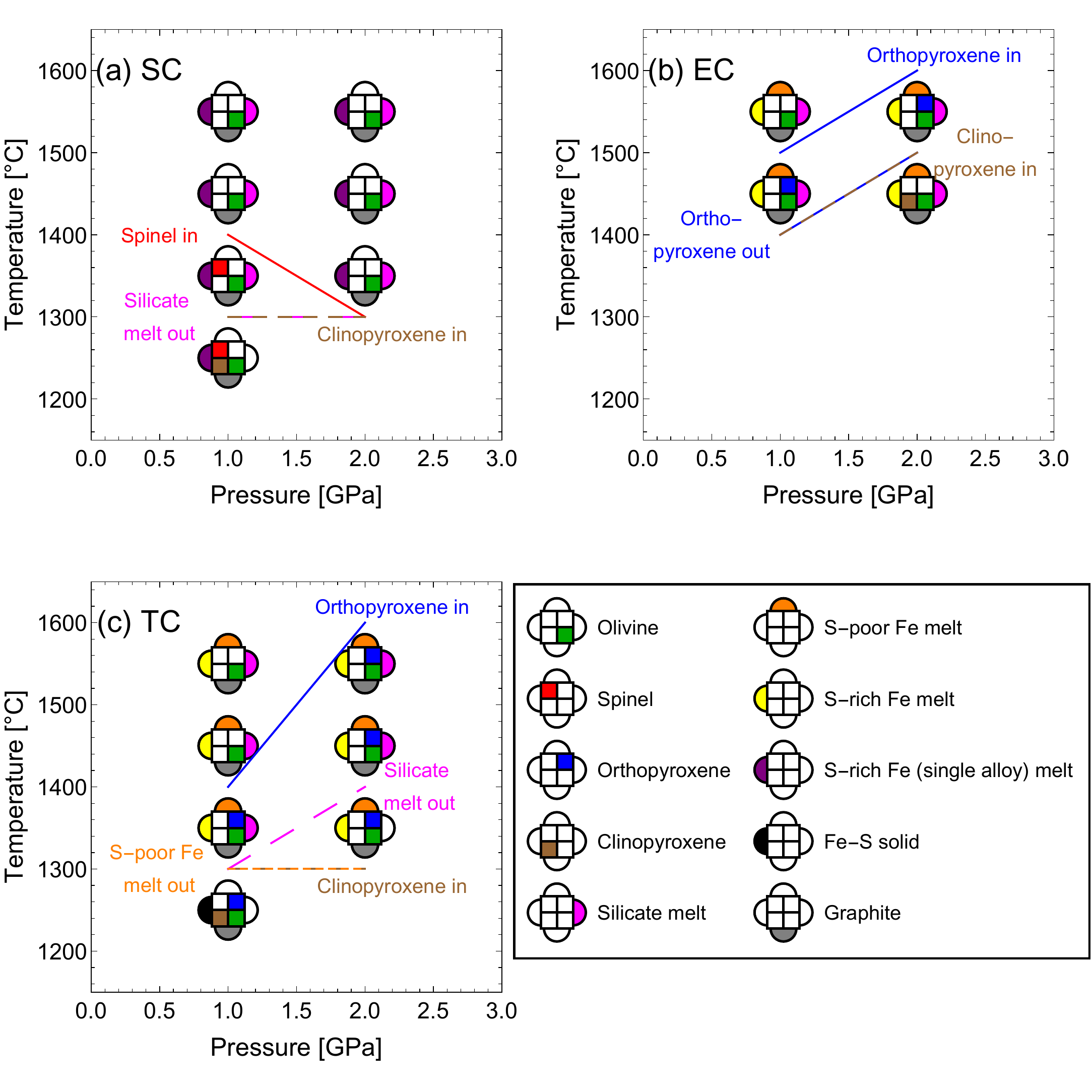}
  \caption[Phase Diagram]{ Phase diagrams of SC (a), EC (b) and TC (c) run products. Solid and dashed lines represent major phase changes in the direction of lower temperatures. }
  \label{fig:PhaseDiagram}
\end{figure*}

Run product phases are listed in Table~\ref{tab:ExperimentConditions}. A clear segregation into silicate and iron-rich phases can be seen in all three series of run products (Fig. \ref{fig:MicroprobeImages}). Resulting phase diagrams for experiments with SC, EC and TC compositions are compared with each other in Fig. \ref{fig:PhaseDiagram}. Oxygen fugacities of EC are lower than those of SC runs by an average value of 0.6 log units (Table~\ref{tab:ExperimentConditions}) since sequential condensation models of \citet{Moriarty2014} are richer in oxygen than equilibrium chemistry models. The oxygen fugacities of carbon-saturated experiments with EC and TC bulk compositions are similar because the relative elemental abundances of equilibrium chemistry models, excluding carbon, for HD19994 and the Sun are largely the same (Table~\ref{tab:StartingMaterial}). Mass-balance calculations on iron alloys and silicate phases, excluding graphite, result in 10$-$18~wt\% of iron alloys in SC runs and 23$-$27~wt\% of iron alloys in EC/TC runs. 

%\afterpage{
%\clearpage
%}

%\afterpage{
%\clearpage
%}

In our run products, graphite grains \textless 1$-$100~$\mathrm{\mu m}$ in diameter (see Fig. \ref{fig:MicroprobeImages}a,b,d) were identified with EDS analyses showing a clear peak of carbon with no other elements. In EC runs with 5 wt\% SiC in their starting material we did not find any SiC grains, suggesting the formation of graphite via the oxidation of silicon in SiC \citep[see][]{Hakim2018b}. Since our experiments were conducted in graphite capsules, all our run products are graphite-saturated, and hence graphite is a stable phase in all runs. 

Olivine crystals are present in all runs. Orthopyroxene is present in all EC runs except the run at 1~GPa and~1823~K, and all TC runs except the runs at 1~GPa and 1723$-$1823~K. The absence of orthopyroxene in SC runs is due to their higher oxygen fugacities and corresponding higher FeO content. Clinopyroxene is present only at 1~GPa and the lowest temperature in all three series. In SC runs at 1~GPa and 1545$-$1623~K, spinel is also identified. Silicate melts and iron alloys are usually concentrated between the boundaries of silicate crystals and at the top or edges of capsules. The proportion of silicate melt increases with temperature and decreases with pressure. The solidus of silicate melt in SC runs is lower than in EC/TC runs due to higher oxygen fugacities and corresponding higher FeO content. 

Iron alloys are present in all runs. In all EC/TC runs except TC run at 1~GPa and 1523~K containing solid Fe-S, two immiscible iron-rich melts (S-rich Fe melt and S-poor Fe melt) are identified. S-poor Fe melt is observed as almost spherical blebs usually surrounded by S-rich Fe melt (see Fig. \ref{fig:MicroprobeImages}b,c,f). This immiscibility is attributed to the chemical interactions between carbon and sulfur. Such liquid metal immiscibility in Fe-C-S systems has been observed for a range of S/Fe ratios in previous studies \citep[e.g.][]{Wang1991,Corgne2008,Dasgupta2009}. The S/Fe ratio in iron-rich melts of our EC/TC runs is within this range (see Sect. \ref{paper2:phaseCompositions}). In contrast, SC runs do not exhibit liquid metal immiscibility and contain a single alloy of S-rich Fe melt, since the S/Fe ratio in this series is beyond the range where immiscibility exists (see below).

As a result of quenching, the silicate melt exhibits a dendritic texture as shown in Fig. \ref{fig:MicroprobeImages}a,b,e,f. Dendrites in SC runs (e.g., Fig. \ref{fig:MicroprobeImages}a,e) are 5$-$10 times larger than in EC/TC runs (e.g., Fig. \ref{fig:MicroprobeImages}b,f). Quenching also results in the growth of thin rims at the boundaries of silicate crystals (e.g., Fig. \ref{fig:MicroprobeImages}a). These thin rims sometimes have a saw-toothed edge and are thicker in SC runs than in EC/TC runs, perhaps due to different viscosities and hence transport properties of the melts involved owing to different oxygen fugacities \citep[e.g.,][]{Giordano2008}. These textures are observed in melt regions because quenching is a non-instantaneous process leading to rapid exsolution and crystallization of melt. The silicate melt also contains Fe-S micro-inclusions resulting from the exsolution of the original melt upon quenching (see Fig. \ref{fig:MicroprobeImages}e), similar to observations made in the study of \citet{Boujibar2014}.

\afterpage{
%\clearpage
%\thispagestyle{empty}
\begin{landscape}
\centering
\begin{table*}[!ht]
\footnotesize
%\small
\caption{\label{tab:PhaseCompositionSilicates} Composition of silicate phases} 
%\tiny
\begin{center}
%\resizebox{\textwidth}{!}{
\renewcommand{\arraystretch}{0.5}
\begin{tabular}{lr|rrrrrcrrr} \hline \hline \rule[2mm]{0mm}{0mm}
Run no. & $n$ & \multicolumn{1}{|c}{$\mathrm{SiO_{2}}$} & \multicolumn{1}{c}{MgO} & \multicolumn{1}{c}{FeO} & \multicolumn{1}{c}{$\mathrm{Al_{2}O_{3}}$} & \multicolumn{1}{c}{CaO} & \multicolumn{1}{c}{S} & \multicolumn{1}{c}{Sum} & \multicolumn{1}{c}{$K_{D}$} & \multicolumn{1}{c}{$K'_{D}$} \\[2mm]																
\hline \\[-1mm]
\multicolumn{3}{@{}l}{\makecell[l]{Olivine}} \\
1B1t  & 19 & 38.5 (0.1) & 39.9 (0.1) & 21.3 (0.1) & 0.1 (0.0) & 0.1 (0.0) & $<$DL & 99.9 (0.1) & 0.26 & 0.26 \\
1B1p  & 10 & 38.2 (0.1) & 40.0 (0.1) & 20.9 (0.2) & 0.1 (0.0) & 0.1 (0.0) & $<$DL & 99.3 (0.1) & 0.32 & 0.28 \\
1B1f  & 12 & 37.6 (0.1) & 37.7 (0.1) & 23.7 (0.1) & 0.1 (0.0) & 0.1 (0.0) & $<$DL & 99.3 (0.1) & 0.27 & 0.26 \\
1B1j  & 10 & 36.5 (0.1) & 34.5 (0.3) & 28.5 (0.5) & 0.1 (0.0) & 0.2 (0.0) & $<$DL & 99.9 (0.3) & 0.30 & 0.26 \\
1B1q  & 10 & 35.3 (0.1) & 29.8 (0.4) & 34.0 (0.1) & 0.2 (0.2) & 0.3 (0.1) & $<$DL & 99.5 (0.2) & 0.25 & 0.27 \\
1B1w  & 10 & 37.0 (0.1) & 34.5 (0.2) & 27.5 (0.2) & 0.1 (0.0) & 0.2 (0.0) & $<$DL & 99.4 (0.1) & 0.25 & 0.25 \\
\vspace{0.1cm}
1B1r  &  9 & 35.4 (0.3) & 28.2 (0.3) & 35.0 (0.1) & 0.6 (0.5) & 0.4 (0.1) & $<$DL & 99.5 (0.3) & $-$ & $-$ \\
2C1a  &  9 & 39.4 (0.1) & 48.0 (0.1) & 12.9 (0.1) & 0.1 (0.0) & 0.1 (0.0) & $<$DL & 100.5 (0.1) & 0.24 & 0.29 \\
2C1d  & 10 & 39.7 (0.1) & 43.7 (0.3) & 17.6 (0.1) & 0.2 (0.2) & 0.2 (0.1) & $<$DL & 100.5 (0.2) & 0.38 & 0.28 \\
2C1e  & 10 & 38.6 (0.1) & 44.7 (0.6) & 16.1 (0.7) & 0.2 (0.2) & 0.2 (0.0) & $<$DL & 99.8 (0.4) & 0.26 & 0.28 \\
\vspace{0.1cm}
2C1c$^\dagger$ & 10 & 38.0 (0.2) & 39.7 (0.2) & 21.5 (0.3) & 0.5 (0.2) & 0.2 (0.0) & $<$DL & 99.9 (0.2) & $-$ & $-$ \\
1A2y  & 10 & 40.3 (0.3) & 44.7 (0.1) & 15.0 (0.1) & 0.0 (0.0) & 0.1 (0.0) & $<$DL & 100.2 (0.2) & 0.31 & 0.28 \\
1A2zc  & 9 & 39.9 (0.1) & 42.8 (0.1) & 16.7 (0.1) & 0.1 (0.0) & 0.1 (0.0) & $<$DL & 99.5 (0.1) & 0.35 & 0.28 \\
1A2zd  & 11 & 39.9 (0.2) & 45.2 (0.5) & 14.3 (0.5) & 0.0 (0.0) & 0.1 (0.0) & $<$DL & 99.6 (0.3) & 0.31 & 0.28 \\
1A2a$^\dagger$  & 8 & 38.4 (0.2) & 41.9 (0.3) & 19.6 (0.2) & 0.1 (0.0) & 0.2 (0.0) & $<$DL & 100.2 (0.2) & $-$ & $-$ \\
1A2c$^\dagger$  & 7 & 37.5 (0.3) & 42.4 (0.5) & 19.5 (0.1) & 0.1 (0.0) & 0.3 (0.0) & $<$DL & 99.9 (0.3) & $-$ & $-$ \\
1A2za  & 17 & 39.7 (0.1) & 40.5 (0.2) & 19.7 (0.1) & 0.1 (0.0) & 0.2 (0.0) & $<$DL & 100.3 (0.1) & $-$ & $-$ \\
1A2s  & 8 & 38.6 (0.1) & 40.7 (0.2) & 20.3 (0.3) & 0.1 (0.0) & 0.2 (0.0) & $<$DL & 99.8 (0.2) & $-$ & $-$ \\

\multicolumn{3}{@{}l}{\makecell[l]{Orthopyroxene}} \\
2C1d  & 8 & 54.8 (0.2) & 30.8 (0.4) & 10.7 (0.4) & 3.0 (0.4) & 0.9 (0.1) & $<$DL & 100.3 (0.3) & 0.32 &  \\
\vspace{0.1cm}
2C1e  & 8 & 56.4 (0.6) & 31.8 (0.2) & 9.5 (0.1) & 2.0 (0.4) & 0.8 (0.0) & $<$DL & 100.6 (0.3) & 0.22 &  \\
1A2zc  & 10 & 56.4 (0.1) & 30.8 (0.1) & 10.9 (0.1) & 0.9 (0.0) & 0.7 (0.0) & $<$DL & 99.8 (0.1) & 0.32 &  \\
1A2a$^\dagger$  & 8 & 54.8 (0.1) & 29.8 (0.5) & 12.2 (0.1) & 2.0 (0.2) & 1.2 (0.1) & $<$DL & 100.0 (0.2) & $-$ &  \\
1A2c$^\dagger$  & 8 & 53.7 (0.9) & 30.4 (0.4) & 12.1 (0.2) & 2.3 (0.1) & 1.3 (0.1) & $<$DL & 99.8 (0.5) & $-$ &  \\
1A2za  & 17 & 55.8 (0.5) & 28.6 (0.4) & 12.4 (0.3) & 2.3 (0.4) & 1.5 (0.2) & $<$DL & 100.7 (0.3) & $-$  \\
1A2s  & 7 & 53.3 (1.2) & 28.9 (0.6) & 12.1 (0.4) & 3.1 (0.5) & 2.1 (0.1) & $<$DL & 99.5 (0.7) & $-$ & \\

\multicolumn{3}{@{}l}{\makecell[l]{Clinopyroxene}} \\
1B1r  & 10 & 48.6 (0.4) & 20.0 (0.4) & 21.1 (0.1) & 7.3 (0.6) & 2.4 (0.4) & $<$DL & 99.4 (0.4) & & \\
2C1c$^\dagger$  & 4 & 51.2 (0.4) & 25.3 (0.4) & 13.8 (0.4) & 7.5 (0.6) & 2.3 (0.1) & $<$DL & 100.0 (0.4) & & \\
1A2s  & 8 & 49.2 (0.5) & 17.8 (1.0) & 7.3 (0.2) & 9.3 (0.9) & 15.8 (1.0) & $<$DL & 99.4 (0.8) & & \\

\multicolumn{3}{@{}l}{\makecell[l]{Spinel}} \\
1B1q  & 8 & 0.3 (0.1) & 13.3 (0.1) & 24.6 (0.2) & 60.9 (0.3) & 0.1 (0.0) & $<$DL & 99.1 (0.2) & & \\
1B1r  & 7 & 0.7 (0.4) & 12.9 (0.2) & 27.3 (0.1) & 58.6 (0.6) & 0.1 (0.0) & $<$DL & 99.6 (0.4) & & \\

\multicolumn{3}{@{}l}{\makecell[l]{Silicate melt}} \\
1B1t  & 9 & 37.0 (0.5) & 17.0 (1.2) & 34.2 (1.1) & 6.1 (0.7) & 3.2 (0.3) & 0.7 (0.2) & 98.1 (0.8) & & \\
1B1p  & 10 & 41.9 (0.4) & 17.0 (0.5) & 27.9 (0.5) & 6.6 (0.3) & 3.2 (0.1) & 0.6 (0.1) & 97.0 (0.3) & & \\
1B1f  & 10 & 39.6 (0.3) & 13.5 (0.4) & 31.2 (0.4) & 7.4 (0.2) & 4.2 (0.1) & 0.8 (0.1) & 96.8 (0.3) & & \\
1B1j  & 8 & 33.5 (0.5) & 14.0 (1.8) & 38.9 (1.5) & 7.2 (1.3) & 3.4 (0.6) & 0.4 (0.0) & 97.4 (1.1) & & \\
1B1q  & 10 & 38.2 (0.3) & 7.0 (0.7) & 32.0 (0.3) & 12.8 (0.3) & 7.1 (0.2) & 0.6 (0.0) & 97.8 (0.4) & & \\
\vspace{0.1cm}
1B1w  & 11 & 33.9 (0.3) & 12.0 (1.4) & 38.5 (0.6) & 8.9 (1.0) & 4.1 (0.5) & 0.8 (0.2) & 98.1 (0.8) & & \\
2C1a  & 6 & 50.0 (1.4) & 15.2 (2.5) & 17.3 (1.4) & 9.2 (1.2) & 5.3 (0.6) & 0.3 (0.1) & 97.2 (1.4) & & \\
2C1d  & 6 & 43.7 (1.9) & 19.7 (3.7) & 21.0 (1.3) & 8.7 (2.1) & 4.9 (1.0) & 0.1 (0.0) & 98.1 (2.0) & & \\
\vspace{0.1cm}
2C1e  & 8 & 48.4 (1.0) & 11.3 (2.5) & 15.7 (1.1) & 13.4 (1.6) & 8.0 (0.8) & 0.2 (0.0) & 97.0 (1.4) & & \\
1A2y  & 14 & 50.6 (0.5) & 18.8 (1.7) & 20.6 (0.8) & 5.2 (0.3) & 4.1 (0.3) & 0.4 (0.1) & 99.7 (0.8) & &  \\
1A2zc & 11 & 45.4 (0.7) & 19.9 (0.4) & 22.1 (0.5) & 6.1 (0.2) & 4.8 (0.2) & 0.3 (0.1) & 98.7 (0.4) & & \\
1A2zd & 10 & 50.8 (0.5) & 19.6 (0.9) & 19.7 (0.4) & 4.8 (0.3) & 3.7 (0.3) & 0.3 (0.1) & 99.1 (0.5) & & \\
[2mm]

\hline
\end{tabular}
%}
\end{center}\caption*{\footnotesize \textit{Note:} All compositions are in wt\% with 1$\sigma$ error given in parentheses. Sulfur in silicate melts is reported as S since oxygen fugacities are much lower than needed to form sulfates \citep{Jugo2005,Jugo2010}. Runs marked $^\dagger$ contain silicate melt in small quantities but could not be measured using EPMA. $n$ is the number of analytical points. DL: detection limit. Pt is $<$DL in all silicate phases. $K_{D}$ is the olivine-silicate melt FeO-MgO exchange coefficient and $K'_{D}$ is the corrected exchange coefficient from \citet{Toplis2005} (see \ref{paper2:appendixEquilibrium} for mineral-melt equilibrium calculations). }
\end{table*}
\end{landscape}
%\clearpage
}

\afterpage{
%\clearpage
%\thispagestyle{empty}
\begin{landscape}
\centering
\begin{table*}[!ht]
\footnotesize
%\small
\caption{\label{tab:PhaseCompositionIronAlloys} Composition of iron-rich phases} 
%\tiny
\begin{center}
%\resizebox{\textwidth}{!}{
\renewcommand{\arraystretch}{0.5}
\begin{tabular}{lc|ccclrlr} \hline \hline \rule[2mm]{0mm}{0mm}
Run no. & $n$/$m$ & \multicolumn{1}{c}{Fe} & \multicolumn{1}{c}{Pt} & \multicolumn{1}{c}{Si} & \multicolumn{1}{c}{C} & \multicolumn{1}{c}{S} & \multicolumn{1}{c}{O} & \multicolumn{1}{c}{Sum}   \\[2mm]																
\hline \\[-1mm]
\multicolumn{4}{@{} l}{\makecell[l]{S-rich Fe melt (single alloy)}} \\
1B1t & 10/4 & 62.7 (0.3) & 0.4 (0.3) & 0.4 (0.5) & 0.7 (0.4) & 29.1 (0.6) & 5.7 (0.8) &  98.9 (0.5) \\
1B1p & 12/5 & 61.1 (0.6) & 1.6 (0.4) & 0.1 (0.1) & 0.5 (0.2) & 33.6 (1.3) & 3.5 (2.9) & 100.5 (0.7)\\
1B1f & 8/10 & 62.6 (0.2) & 0.3 (0.2) & 0.1 (0.1) & 0.4 (0.2) & 30.3 (0.9) & 5.1 (0.8) &  98.8 (0.5) \\
1B1j &  9/9 & 63.0 (0.8) & 1.8 (0.4) & 0.2 (0.1) & 0.5 (0.2) & 29.4 (1.2) & 6.5 (1.0) & 101.3 (0.7)\\
1B1q &  8/8 & 64.0 (0.3) & 0.4 (0.1) & 0.1 (0.1) & 0.2 (0.1) & 28.5 (1.7) & 6.0 (1.6) &  99.3 (1.0)\\
1B1w & 9/10 & 63.4 (0.2) & 0.1 (0.1) & 0.3 (0.1) & 0.4 (0.2) & 29.4 (0.9) & 6.7 (0.6) & 100.3 (0.5)\\
\vspace{0.1cm}
1B1r &  8/9 & 61.9 (1.0) & 0.9 (0.8) & $<$DL     & 0.3 (0.1) & 37.3 (0.2) & 0.7 (0.2) & 101.1 (0.5) \\

\multicolumn{3}{@{} l}{\makecell[l]{S-rich Fe melt}} \\
2C1a  & 5/10 & 69.2 (0.3) & $<$DL     & $<$DL & 1.0 (0.7)$^\dagger$ & 30.3 (0.2) & 0.9 (0.3)$^\dagger$ & 101.4 (0.2) \\
2C1d  & 5/10 & 70.3 (1.4) & 0.2 (0.1) & $<$DL & 1.0 (0.7) & 27.0 (0.4) & 0.9 (0.3) & 99.4 (0.6)\\
2C1e  &  5/4 & 68.8 (1.9) & 0.1 (0.1) & $<$DL & 1.6 (0.7) & 30.2 (1.0) & 1.3 (0.8) & 102.0 (1.0) \\
2C1c  & 5/4  & 69.4 (1.0) & 0.3 (0.1) & $<$DL & 1.6 (0.7)$^\dagger$ & 28.9 (1.2) & 1.3 (0.8)$^\dagger$ & 101.6 (0.8) \\
1A2y & 10/10 & 68.4 (0.3) & $<$DL     & $<$DL & 0.4 (0.2) & 29.8 (0.9) & 1.0 (0.5) & 99.6 (0.4) \\
1A2zc &  7/7 & 68.6 (0.8) & 0.3 (0.1) & $<$DL & 0.6 (0.1) & 29.7 (1.6) & 1.1 (0.1) & 100.3 (0.7)\\
1A2zd &15/11 & 69.2 (0.9) & 0.3 (0.0) & $<$DL & 0.4 (0.1) & 30.2 (1.2) & 0.6 (0.2) & 101.6 (0.6)\\
1A2a  & 14/5 & 69.3 (1.2) & 0.2 (0.1) & $<$DL & 0.4 (0.1) & 29.7 (1.4) & 0.9 (0.4) & 100.7 (0.8) \\
1A2c  & 8/10 & 68.9 (0.9) & 0.1 (0.0) & $<$DL & 0.6 (0.2) & 30.2 (1.1) & 0.6 (0.2) & 100.4 (0.6) \\
\vspace{0.1cm}
1A2za &  8/5 & 69.4 (0.6) & 0.3 (0.1) & $<$DL & 0.8 (0.5) & 29.4 (0.6) & 1.1 (0.2) & 100.9 (0.4) \\

\multicolumn{3}{@{} l}{\makecell[l]{S-poor Fe melt}} \\
2C1a  & 5/20 & 94.0 (0.5) & 0.5 (0.1) & $<$DL & 3.6 (0.2) & 0.9 (0.1) & 0.4 (0.0) & 99.4 (0.2) \\
2C1d  & 5/16 & 89.0 (0.5) & 3.5 (0.1) & $<$DL & 3.2 (0.2) & 1.9 (0.1) & 0.4 (0.0) & 98.0 (0.2) \\
2C1e  &  4/5 & 93.8 (0.5) & 0.2 (0.0) & $<$DL & 5.4 (1.0) & 1.7 (0.3) & 0.4 (0.0) & 101.4 (0.5)\\
2C1c  & 4/12 & 88.3 (0.2) & 4.4 (0.2) & $<$DL & 4.5 (0.3) & 1.1 (0.1) & 0.2 (0.0) & 98.5 (0.2) \\
1A2y  & 13/10 & 90.9 (0.4) & 3.6 (0.4) & $<$DL& 2.1 (0.1) & 1.0 (0.1) & 0.3 (0.0) & 97.8 (0.2) \\
1A2zc & 13/9 & 87.4 (0.6) & 6.4 (0.4) & $<$DL & 3.6 (0.6) & 1.4 (0.2) & 0.3 (0.0) & 99.2 (0.4) \\
1A2zd & 9/11 & 87.1 (0.3) & 6.4 (0.2) & $<$DL & 3.3 (0.2) & 1.0 (0.2) & 0.2 (0.0) & 98.1 (0.2) \\
1A2a  & 7/10 & 88.2 (0.7) & 6.4 (0.5) & $<$DL & 4.3 (0.4) & 1.3 (0.2) & 0.3 (0.0) & 100.4 (0.4)\\
1A2c & 10/10 & 89.1 (0.8) & 5.1 (0.8) & $<$DL & 4.5 (0.5) & 1.0 (0.2) & 0.3 (0.1) & 100.0 (0.5) \\
\vspace{0.1cm}
1A2za& 17/13 & 86.2 (0.5) & 8.3 (0.3) & $<$DL & 4.5 (0.9) & 1.6 (0.2) & 0.2 (0.1) & 100.9 (0.4)\\

\multicolumn{4}{@{} l}{\makecell[l]{Fe-S solid}} \\
1A2s  & 9/5 & 61.3 (0.1) & 0.2 (0.1) & $<$DL & 0.5 (0.2) & 38.9 (0.1) & 0.3 (0.1) & 101.1 (0.1) \\[2mm]

\hline
\end{tabular}
%}
\end{center}\caption*{\footnotesize \textit{Note:} All compositions are in wt\% with 1$\sigma$ error given in parentheses. $n$ is the number of analytical points for Fe and Pt, and $m$ is the number of analytical points for other elements. DL means below detection limit. Mg, Ca and Al are $<$DL in all iron-rich phases. The Pt contamination is 0$-$2.2~mol\% in S-poor Fe melt, negligible in S-rich Fe melt and 0$-$0.4~mol\% in S-rich Fe melt (single alloy). The numbers marked with $^\dagger$ were not measured for that phase and have been taken from the same phase of another run product at similar conditions. }
\end{table*}
\end{landscape}
%\clearpage
}

The iron-rich melts show a fine-grained quench texture supporting the interpretation of a liquid state during the experiments. In EC/TC runs, the immiscibility of S-poor and S-rich Fe melts is evident from the sharp boundaries between them (see Fig. \ref{fig:MicroprobeImages}a,c,f). Sub-micron sized iron-rich specks seen in S-rich Fe melt, surrounding the S-poor Fe blebs, are likely a result of unmixing upon quenching.

\subsection{Phase compositions}\label{paper2:phaseCompositions}

Tables \ref{tab:PhaseCompositionSilicates} and \ref{tab:PhaseCompositionIronAlloys} list the compositions of silicate and iron-rich phases, respectively. The lithophile elements, Mg, Si, Al and Ca are bonded to oxygen in silicate phases. O is largely present in silicate phases and to a smaller extent in iron-rich melts. S mainly partitions into iron-rich phases with smaller amounts present in silicate melts. Fe is distributed among silicate and iron-rich phases. Most of the carbon is present as graphite and a smaller amount is present in iron alloys. 

Olivine crystals and silicate melts in EC/TC runs are richer in MgO and poorer in FeO than in SC runs. The Mg\# , or Mg/(Mg+Fe) mol\% of olivine in EC/TC and SC runs is between 75$-$87 and 55$-$75, respectively. Similarly, the Mg\# of silicate melt in EC/TC and SC runs is between 60$-$65 and 25$-$50, respectively. Orthopyroxene, found only in certain EC/TC runs, has Mg\# between 80$-$87. The SiO$_{2}$ content of silicate melt in EC/TC runs (44$-$50~wt\%) is higher than in SC runs (33$-$41~wt\%). These differences between EC/TC and SC runs are a direct consequence of lower oxygen fugacities of EC/TC runs ($\log f_{\mathrm{O_{2}}}\sim$IW$-$1.1) compared to SC runs ($\log f_{\mathrm{O_{2}}}\sim$IW$-$0.5) (see Table~\ref{tab:ExperimentConditions}). The MgO content of olivine, orthopyroxene and silicate melt increases and the FeO content decreases with temperature.

Since olivines do not accommodate significant amounts of the oxides of Ca and Al, they are present only in silicate melt and/or pyroxenes. Orthopyroxenes and clinopyroxenes contain a combined 2$-$5~wt\% and 10$-$25~wt\% of CaO and $\mathrm{Al_{2}O_{3}}$, respectively. The similarity in CaO and $\mathrm{Al_{2}O_{3}}$ contents of clinopyroxenes between SC and EC runs and their differences from TC runs are due to the differences in starting Ca/Si and Al/Si ratios. Silicate melts contain 0.1$-$0.8 wt\% S, with higher values of S seen mainly in SC runs, which is likely due to their higher FeO contents than in the EC/TC runs \citep[e.g.,][]{Smythe2017}. The formation of spinel in the SC run at 1~GPa and 1545$-$1623~K and its absence in TC runs is likely due to a higher Al/Si ratio of the SC composition.

Across EC/TC runs exhibiting liquid metal immiscibility, the S-poor Fe melt contains 86$-$94~wt\%~Fe, $3.9\pm0.9$~wt\%~C, $1.3\pm0.3$~wt\%~S and $0.3\pm0.1$~wt\%~O. The variable Fe content is a result of the variable Pt contamination of 0$-$8~wt\% (0$-$2~mol\%) from the outer Pt capsules surrounding the inner graphite capsules. The S-rich Fe melt contains $69.2\pm0.5$~wt\%~Fe, $0.8\pm0.5$~wt\%~C, $29.5\pm1.1$~wt\%~S and $1.0\pm0.3$~wt\%~O. The S-rich Fe melt (single alloy) in SC runs contains $62.7\pm0.9$~wt\%~Fe, $0.4\pm0.1$~wt\%~C, 29$-$37~wt\%~S and 1$-$7~wt\%~O ($\sim$36~wt\% S$+$O, equivalent to sulfur's composition in iron sulfide). The higher amount of oxygen in SC iron alloys is likely due to their higher oxygen fugacity with respect to EC/TC iron alloys.

\begin{figure*}[!ht]
  \centering
  \medskip
  \includegraphics[width=.7\textwidth]{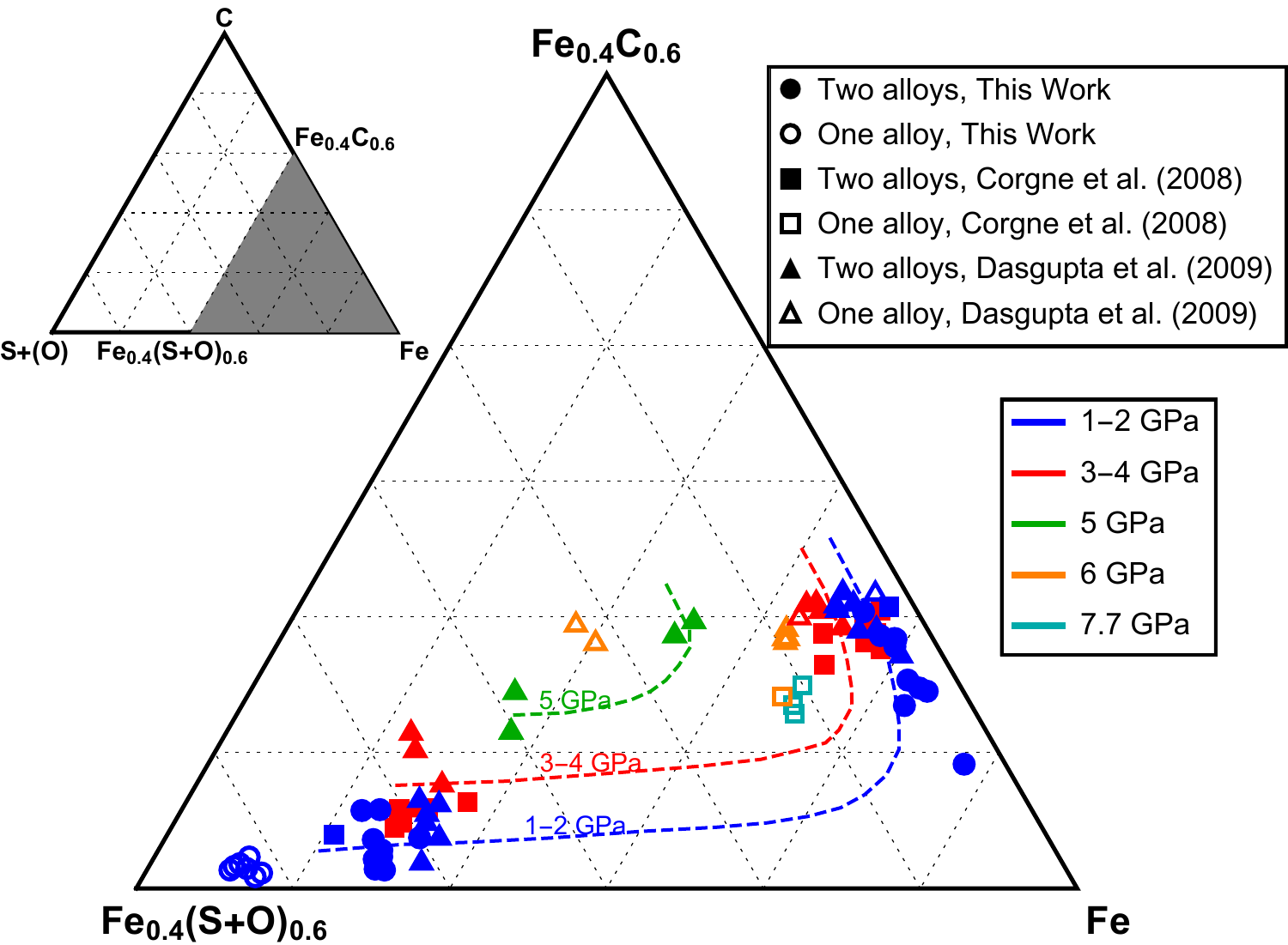}
  \caption[Fe-C-S Ternary Diagram]{Liquid metal immiscibility in the Fe-C-S system compared with results from previous studies. For all studies, O measurements are added to S. For \citet{Corgne2008}, Ni measurements are added to Fe. The hand-drawn dashed lines based on the experiments considered here represent the compositional variation of immiscible S-poor and S-rich Fe melts with pressure.}
  \label{fig:FeSCTernaryDiagram}
\end{figure*}

Fig.~\ref{fig:FeSCTernaryDiagram} illustrates that our measurements of S-rich and S-poor Fe melts exhibiting immiscibility are in excellent agreement with the studies of \citet{Corgne2008} and \citet{Dasgupta2009}. The single alloys from our SC runs are clustered together in the lower left corner of Fig. \ref{fig:FeSCTernaryDiagram} and their composition is a reflection of the starting composition, as is the case for single alloys reported in \citet{Corgne2008} and \citet{Dasgupta2009}. The molar S/Fe ratio in bulk iron-rich melts of our SC runs is $\sim$0.85, which is higher than that of our EC and TC runs having $\sim$0.4 and $\sim$0.25, respectively. Up to pressures of 4$-$6~GPa, \citet{Dasgupta2009} observed immiscibility for S/Fe ratios of $\sim$0.1 and $\sim$0.33 and miscibility for S/Fe ratios of 0.02 and 0.06. \citet{Corgne2008} also found immiscibility at S/Fe$\sim$0.15. Since the miscibility gap closes above 4$-$6~GPa, some runs contain single alloys despite having characteristic S/Fe ratios. Combined with our results this implies that immiscibility is observed in the Fe-C-S system for moderate S/Fe ratios between $\sim$0.1$-$0.8 up to pressures of 4$-$6~GPa. For lower or higher S/Fe ratios, a single iron-rich melt is expected.

\section{Mineralogy and structure of C-enriched rocky exoplanets}\label{paper2:interior}

\subsection{Mineralogy}\label{paper2:mineralogy}

Although our experimental conditions are valid for the interior of Pluto-mass planets as the shallow upper mantles of larger planets, here we discuss mineralogy in the context of both smaller and larger planets. Our experiments show that silicate minerals, iron-rich alloys and graphite dominate the mineralogy in differentiated C-enriched planetary interiors. In addition to the C/O ratio, the oxygen fugacity and the Mg/Si, Al/Si, Ca/Si and S/Fe ratios play an important role in determining the mineralogy in carbon-rich conditions. Our oxygen fugacity conditions (IW$-$0.3$< \log f_{\mathrm{O_{2}}}<$IW$-$1.2) are a direct reflection of the chemical modeling calculations of \citet{Moriarty2014} who calculated the bulk composition of planetesimals in protoplanetary disks. However, oxygen fugacity can vary over a larger range either because of its dependence on pressure implying its change with depth in planetary interiors, or due to the formation in protoplanetary disks of more reducing bulk compositions than the ones we used in this study. Here we place our results in a broader context of C-enriched planetary interiors by combining our results with previous studies.

At IW$< \log f_{\mathrm{O_{2}}}<$IW$-$2 and up to $\sim$4$-$6~GPa, our study and previous studies \citep[e.g.,][]{Dasgupta2009,Dasgupta2013b} show that iron-rich melts are composed of single alloys or two immiscible alloys in the Fe-C-S system depending on the S/Fe ratio. The immiscible S-poor and S-rich~Fe melts show characteristic solubilities with $\sim$5~wt\%~C and $\sim$1~wt\%~S, and $\sim$1~wt\%~C and $\sim$30~wt\%~S, respectively. At pressures higher than $\sim$4$-$6~GPa, a closure of the miscibility gap will allow only single Fe-C-S alloys \citep{Corgne2008,Dasgupta2009}. With decreasing $\log f_{\mathrm{O_{2}}}$ from IW$-$2 to IW$-$6, the solubility of C in Fe decreases and the solubility of Si in Fe increases \citep{Deng2013,Li2016}. At even lower oxygen fugacities ($\log f_{\mathrm{O_{2}}}\sim$IW$-$6.2), C is not soluble in Fe and about $\sim$20~wt\%~Si is present \citep{Takahashi2013}. \citet{Morard2010} show that the Fe-S-Si system also exhibits liquid metal immiscibility similar to the Fe-C-S system at $\log f_{\mathrm{O_{2}}}\sim$IW$-$10, although this miscibility gap closes at approximately 25~GPa. If the temperature is lower than the liquidus in the Fe-C{$\pm$}S{$\pm$}Si system, solids such as Fe, FeS, $\mathrm{Fe_{3}C}$, $\mathrm{Fe_{7}C_{3}}$ and Fe-Si can form depending on pressure and $f_{\mathrm{O_{2}}}$ \citep{Deng2013}.

\citet{Takahashi2013} found olivine to be a dominant silicate mineral in the FMS+CO system at IW$-$1$< \log f_{\mathrm{O_{2}}}<$IW$-$3.3 and 4~GPa. Our experiments in the FCMAS+CSO system at IW$-$0.3$<\log f_{\mathrm{O_{2}}}<$IW$-$1.2 and 1$-$2~GPa showed a larger variety in silicate minerals such as orthopyroxene, clinopyroxene and spinel, in addition to olivine. The diversity in silicates increases with decreasing temperature. The compositions of these silicate minerals are sensitive to $f_{\mathrm{O_{2}}}$ and Mg/Si, Ca/Si and Al/Si ratios. At their lowest $\log f_{\mathrm{O_{2}}}\sim$IW$-$6.2, \citet{Takahashi2013} find periclase to be a dominant mineral because of decreased concentration of $\mathrm{SiO_{2}}$. At pressures above 25~GPa, olivine polymorphs break down to form perovskite and ferropericlase \citep{HiroseFei2002}.

\begin{figure*}[!ht]
  \centering
  \medskip
  \includegraphics[width=.5\textwidth]{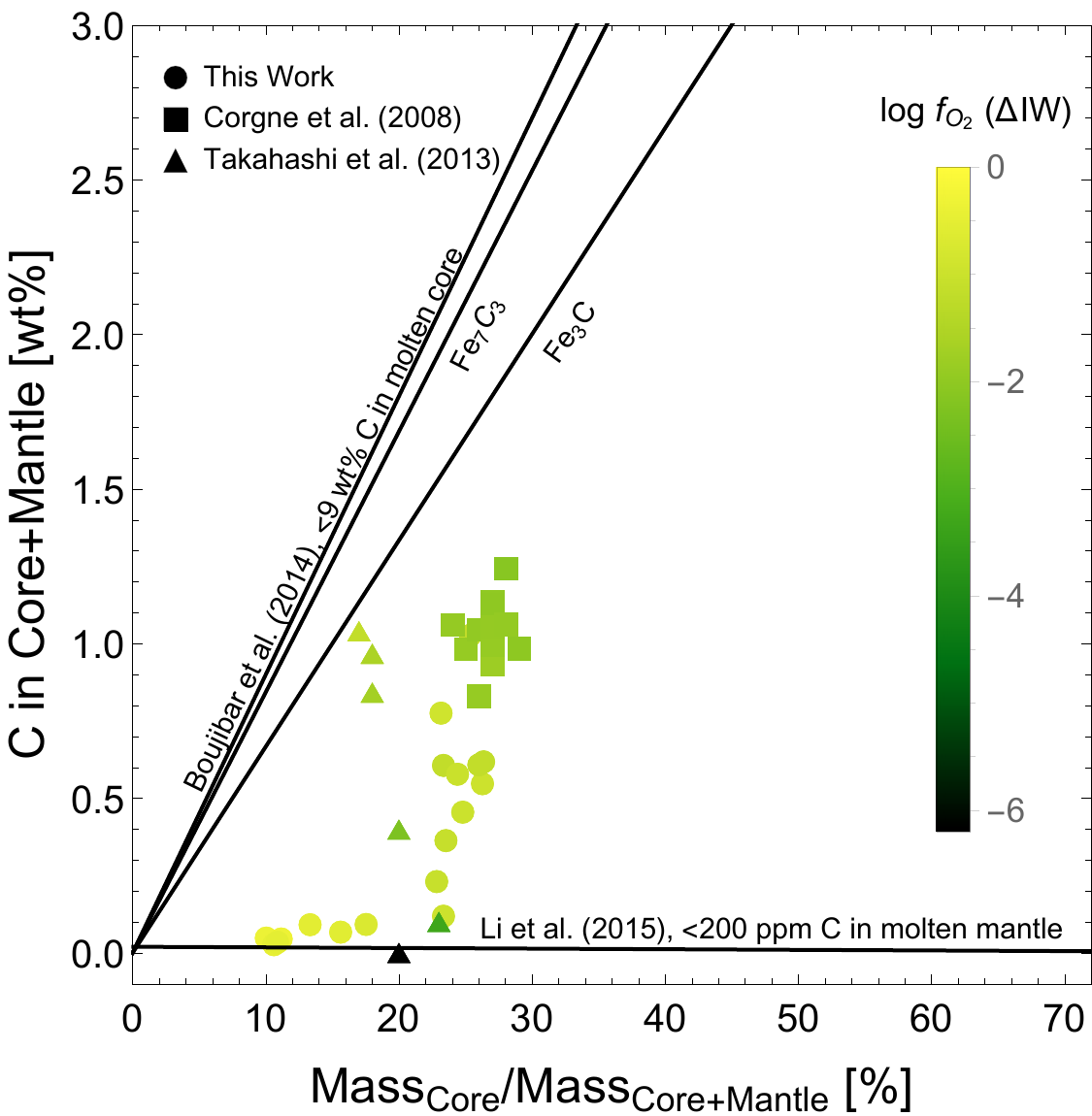}
  \caption[Carbon in Core and Mantle]{Experimental measurements of carbon solubility in iron alloys is plotted against the core mass per cent. Solid lines give upper bounds on carbon solubility in a molten iron-rich core and a molten silicate mantle. }
  \label{fig:CarbonInCore}
\end{figure*}

We do not observe any carbonates in our runs since magnesite and calcite are stable only at very oxidizing conditions, $\log f_{\mathrm{O_{2}}}>$IW$+$1 \citep{Rohrbach2011,Lazar2014}. We also do not find silicon carbide in our runs, as it forms only at extremely reducing conditions, $\log f_{\mathrm{O_{2}}}\sim$IW$-$6.2 \citep{Takahashi2013}. Only highly reduced planets containing no oxidized iron (Fe$^{2+}$ or Fe$^{3+}$) can stabilize silicon carbide in their magma ocean stage \citep[see][]{Hakim2018b}. Our starting compositions do not present such extremes in terms of oxygen fugacity and these minerals are therefore unlikely to form in the magma ocean stage of the type of planets we have considered.

Carbon solubility in the interior of the Earth is key in driving the terrestrial carbon cycle \citep{Dasgupta2013a}. Similarly, carbon solubility is expected to impact the carbon cycles and habitability on C-enriched rocky exoplanets \citep{Unterborn2014}. Although we do not measure the carbon abundance in silicate melts of our experiments, \citet{Li2015,Li2016} give an upper limit of $\sim$200~ppm C in silicate melts at oxygen fugacity conditions similar to our experiments. The C-solubility in S-poor Fe melts is 3$-$9~wt\% \citep[][and this study]{Rohrbach2014,Boujibar2014,Li2015,Li2016}, which is more than two orders of magnitude larger than the C-solubility in silicates. 

In Fig.~\ref{fig:CarbonInCore}, we plot the C-solubility in the mantle and core against core mass per cent (excluding graphite) for our experiments and those from \citet{Corgne2008} and \citet{Takahashi2013}. Since all experiments are carbon-saturated, this figure essentially gives the minimum amount of carbon in a planet that is necessary to achieve carbon-saturation in the planet during its magma ocean stage. For a C-enriched exoplanet with a EC/TC-like composition and 25\% of its mass in the core, $\sim$0.7~wt\%~C (molar C/O$\sim$0.03) is sufficient for carbon-saturation. For a C-enriched exoplanet with a SC-like composition and 10\% of its mass in the core, $\sim$0.05~wt\%~C (C/O$\sim$0.002) is sufficient for carbon-saturation. For an extreme case with a zero core mass, the minimum amount of carbon needed for carbon-saturation is 200~ppm (C/O$\sim$0.001). In contrast, if the core mass per cent is Mercury-like (70\%), assuming 9~wt\% C in the core, 6~wt\% C (C/O$\sim$0.5) is needed for carbon-saturation. Once carbon-saturation is achieved, an increase in C/O ratio increases only the amount of graphite produced, and has a negligible impact on the mineralogy of silicates or iron-alloys. The solubilities of carbon for the experiments considered in Fig.~\ref{fig:CarbonInCore} is lower than the 9~wt\% C-solubility from \citet{Boujibar2014} because of the difference in oxygen fugacities and/or the presence of two Fe alloys where S-rich Fe melts have lower C-solubility than S-poor Fe melts, which decreases the net C-solubility in the iron-rich core. The C-solubility in the core is more or less the same for $\log f_{\mathrm{O_{2}}}$ from IW to IW$-$2, whereas it decreases with a further decrease in $\log f_{\mathrm{O_{2}}}$ from IW$-$2 to IW$-$6.2, where it becomes negligible. 

\subsection{Interior structure}\label{paper2:structure}

High temperatures during planet formation enable melting and chemical segregation of several minerals \citep{ElkinsTanton2012}. These minerals eventually undergo gravitational stratification. For C-enriched rocky exoplanets, iron alloys, silicates and graphite are the main categories of minerals based on densities. Due to density contrasts of more than 40\% between graphite and silicates and more than 50\% between silicates and iron alloys, three major gravitationally stable layers are expected to form in these exoplanets: an iron-rich core, a silicate mantle and a graphite layer on top of the silicate mantle.

For smaller C-enriched rocky exoplanets (with interior pressures $<4$~GPa) showing Fe-C-S liquid metal immiscibility, S-poor Fe alloy will form the inner core and S-rich Fe alloy will form the outer core because of the density contrasts between the two alloys. Our mass-balance calculations show that the core/mantle mass ratio would be $\sim$0.33 for planets with EC/TC compositions and $\sim$0.15 for planets with SC composition. Even though the core/mantle ratio is similar for EC and TC compositions, the S-poor Fe inner-core to the S-rich Fe outer-core mass ratio would be about 0.7 for TC planets and about 1.6 for EC planets owing to a difference in the S/Fe ratio. For C-enriched exoplanets with core pressures larger than 6~GPa, there would be no stratification in the core because of the closure of the miscibility gap. For planets with extremely reducing cores showing Fe-S-Si immiscibility, again an inner and an outer core would exist \citep[e.g.,][]{Morard2010}. Depending on composition, pressure, temperature and $f_{\mathrm{O_{2}}}$, cores may stratify into multiple metal-rich layers, for instance, a solid $\mathrm{Fe_{3}C}$ inner core with a liquid S-rich Fe outer core, or a solid Fe inner core surrounded by a solid $\mathrm{Fe_{3}C}$ middle core and a liquid FeS outer core \citep[e.g.,][]{Deng2013}.

Even though C-enriched rocky exoplanets are expected to contain large amounts of carbon, olivine and pyroxenes would be the common mantle minerals, similar to C-poor rocky planets. Additionally, minerals such as spinel and garnet may be abundant in the upper mantle for planets depending on Al/Si and Ca/Si ratios, which might also vary as shown by the models of \citet{CarterBond2012a}. For larger C-enriched rocky exoplanets, high-pressure phases of these minerals, ferropericlase and perovskite and/or post-perovskite would be the most abundant minerals in the lower mantle. 

Graphite will likely form a flotation layer on top of the magma ocean or silicates such as olivine because of its lower density. Graphite is expected to be in its solid state because the melting temperature of the graphite-diamond system exceeds 4500~K for all pressures in a planetary interior \citep{Ghiringhelli2005}. If the graphite layer extends deep into the planet exceeding pressures of 2$-$15~GPa and depending on the temperature, diamond would form beneath graphite. Since diamond is denser than graphite with a density comparable to some silicate minerals, convection, if it exists, in the mantle may strip off diamonds from beneath the graphite layer. This would result in a diamond-silicate mantle similar to the mantle discussed by \citet{Unterborn2014}. Additionally, the possible presence of metastable states in the carbon system, at conditions near the equilibrium graphite-diamond transition, may have interesting consequences for planetary evolution because of their substantially different physical properties compared to those of graphite and diamond \citep[e.g.,][]{Shabalin2014}. However, the discussion of these metastable states is beyond the scope of this study.

\section{A C-enriched interior for Kepler-37b}\label{paper2:kepler37b}

\subsection{Effect of a graphite layer on the derived mass}\label{paper2:kepler37bMass}

Transit photometry is used to measure the radius of exoplanets \citep{Batalha2014}. Follow-up stellar radial velocity measurements help to put constraints on their masses, but for most of the rocky exoplanets, masses are currently unknown. Due to graphite's significantly lower density compared to silicate minerals and iron-rich alloys, the mass of an exoplanet in the presence of significant amounts of graphite would be lower than expected for a given radius. To quantify the effect of graphite on a planet's mass, we compute the interior structure and mass of the smallest known exoplanet till date, Kepler-37b with radius of 0.34 $\mathrm{R_{\Earth}}$ \citep{Stassun2017}, by following the isothermal recipe to solve the hydrostatic and Poisson's gravitational gradient equations and keeping the radius fixed \citep[e.g.,][]{Unterborn2016}. We implement the third-order Birch-Murnaghan equation of state  in order to provide a relation between density and pressure \citep{Birch1947}. Since we are interested in the effect of graphite on its total mass, we assume Kepler-37b is fully differentiated with a pure iron or iron sulfide core, an enstatite mantle and a graphite layer. To model the equations of state, we use the thermoelastic data of graphite \citep{Colonna2011}, enstatite \citep{Stixrude2005}, iron \citep{Fei2016} and iron sulfide \citep{Sata2010}.

\begin{figure*}[!ht]
  \centering
  \begin{subfigure}[b]{0.75\textwidth}
   \includegraphics[width=1\linewidth]{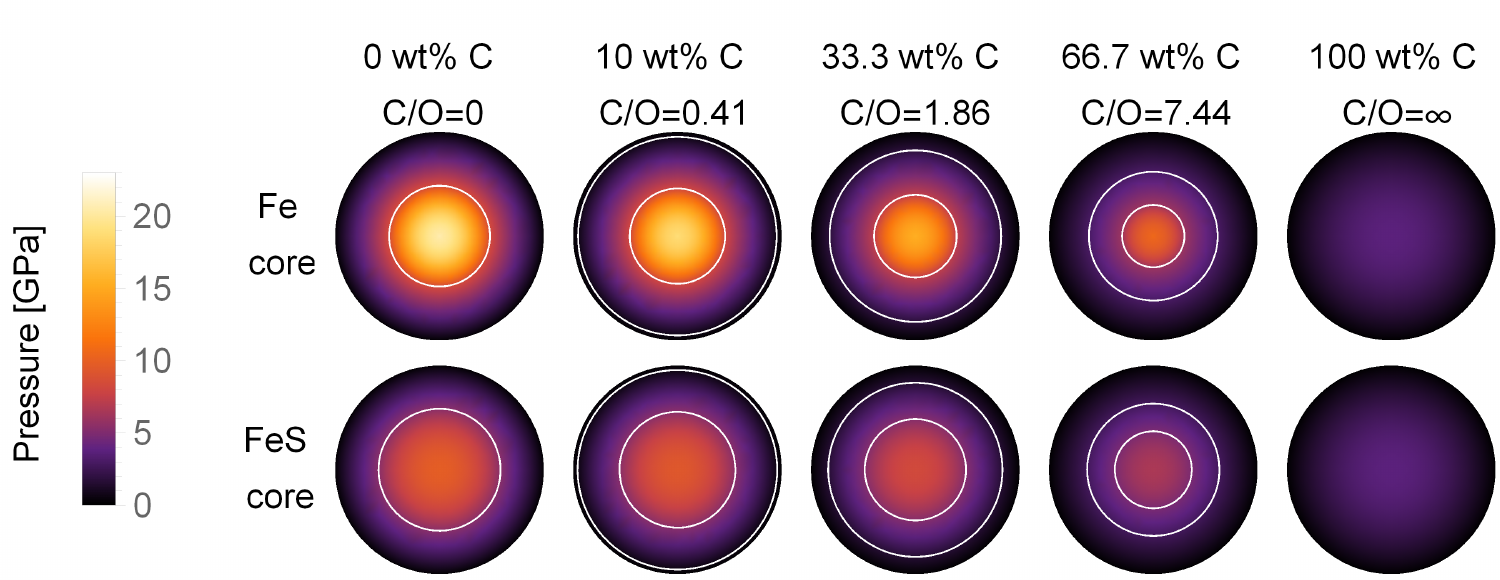}
   \caption{}
   \label{fig:Kepler37bPressure} 
  \end{subfigure}

  \begin{subfigure}[b]{0.75\textwidth}
   \includegraphics[width=1\linewidth]{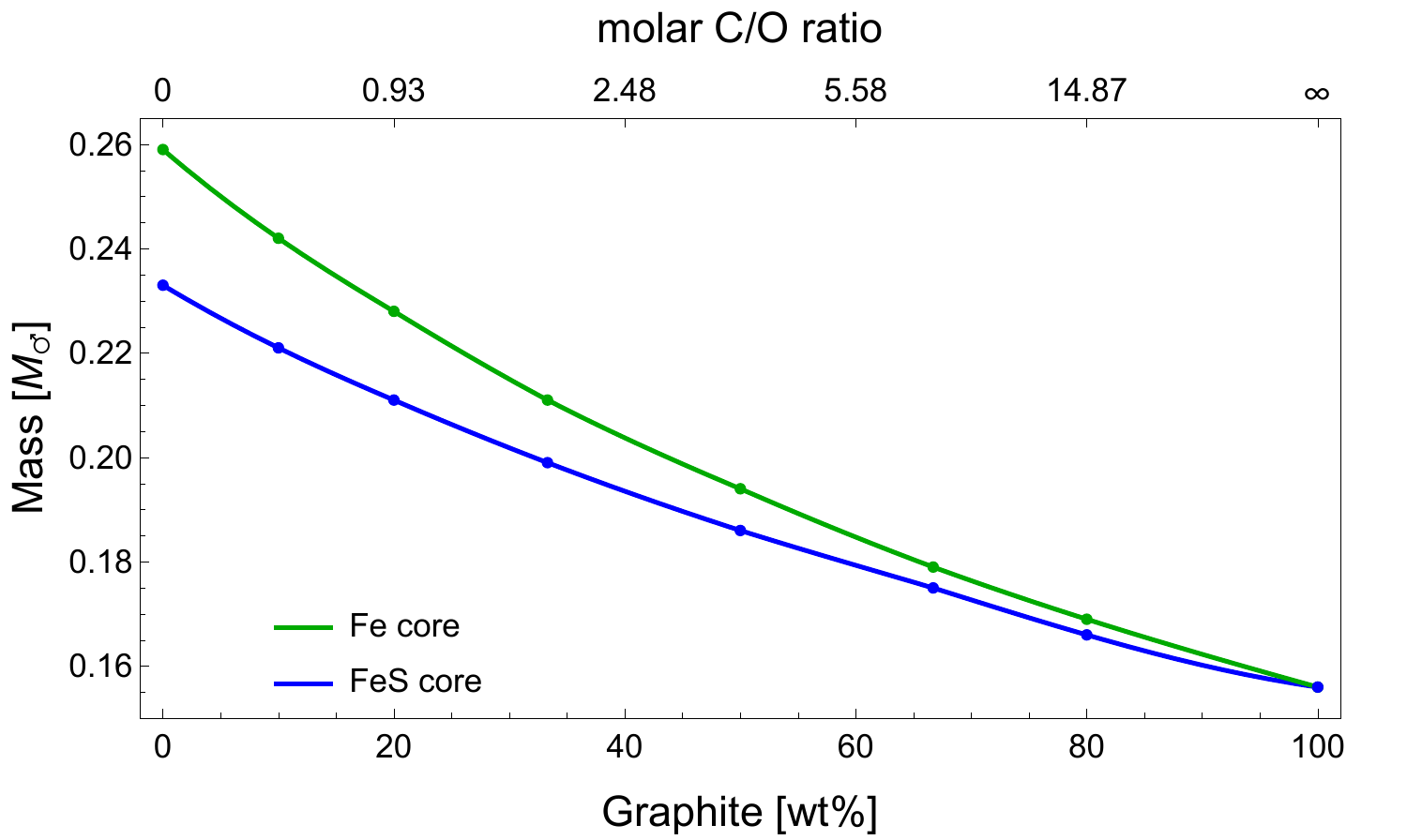}
   \caption{}
   \label{fig:Kepler37bMass}
  \end{subfigure}
\caption[Interior Kepler 37b]{(Bottom) Solid lines represent the derived mass of Kepler-37b assuming a pure iron or an iron sulfide core, an enstatite mantle and a graphite layer with a core/mantle mass ratio of 0.33 and different mass fractions of graphite for a fixed planet's radius of 0.34 $\mathrm{R_{\Earth}}$ \citep{Stassun2017}. (Top) The internal pressure distribution of Kepler-37b is also shown for five cases where white contours represent the core-mantle and mantle-crust boundaries. }
\end{figure*}\label{fig:Kepler37b}

%\afterpage{
%\clearpage
%}

Applying a core/mantle mass ratio of 0.33, similar to our EC/TC results, and assuming a pure Fe core and an enstatite mantle to the interior structure model of Kepler-37b, the derived mass is 0.26 times the martian mass (0.26~$\mathrm{M_{\mars}}$). When a 33.3~wt\% graphite layer is assumed on top of its mantle keeping the core/mantle mass ratio at 0.33, the total mass becomes 0.21 $\mathrm{M_{\mars}}$ (about 19\% less). In fact, a graphite layer of 10~wt\% of the planet's total mass is sufficient to decrease the derived mass of Kepler-102b by 7\% (see Fig.~5, bottom). Assuming a hypothetical 100~wt\% graphite-only planet gives a mass of 0.16~$\mathrm{M_{\mars}}$ (about 40\% less). For models with a FeS core instead of pure Fe, a similar trend can be seen (Fig.~5,bottom). 

With future missions such as TESS \citep{Ricker2014}, CHEOPS \citep{Fortier2014} and PLATO \citep{Ragazzoni2016}, the masses and radii of rocky exoplanets will be measured with higher accuracy. Along with improved knowledge of stellar chemistry, tighter constraints on planetary bulk compositions will also be feasible \citep{Dorn2015,Santos2017}. This in turn will enable better constraints on the presence of low-density minerals like graphite in the interior of rocky exoplanets. 

We also show the internal pressure distribution of Kepler-37b in Fig.~5 (top) for the cases of 0, 10, 33.3, 66.7 and 100 wt\% graphite. The central pressure of Kepler-37b decreases with the amount of graphite. For our models of Kepler-37b, pressures at the bottom of graphite layers are \textless 4~GPa, making phase transformation to diamond impossible at temperatures above 1000~K \citep{Ghiringhelli2005}. C-enriched rocky exoplanets larger than Kepler-37b with thick graphite layers are likely to form diamonds beneath these graphite layers. If the amount of diamond is significantly larger than that of graphite, the effect on the derived mass of the planet would be smaller since the density of diamond is higher than graphite and comparable to silicates. 

\subsection{Observations and habitability}\label{paper2:kepler37bObservations}

The abundance of graphite on the planetary surface will have major consequences for planetary thermal evolution, volatile cycles and atmospheric composition, surface geochemistry and habitability. Identification of such a planet by future observations would be of great significance. Graphite has a very low reflectance compared to usual silicate-rich minerals forming the surface of terrestrial planets such as Mars. If Earth or an exoplanet is covered with a graphite layer, the planet's surface would likely appear to be dark with an albedo much lower than expected for a C-poor planet. Similarly, a darkening agent discovered on Mercury's surface has been speculated to be graphite \citep{Peplowski2016}.

Small C-enriched exoplanets are unlikely to retain a primary atmosphere. Secondary atmospheres of graphite-layered planets might be non-existent if the graphite layers are able to completely isolate the silicate mantles. For planets with relatively thin graphite layers, outgassing processes from the silicate mantle may allow for an atmosphere to exist. Atmospheres of C-enriched rocky exoplanets are believed to be devoid of oxygen-rich gases \citep[e.g.,][]{Kuchner2005}. Carbon is dissolved in silicate melts mainly as CO$_2$ at $\log f_{\mathrm{O_{2}}}>$IW$-$1 and mainly as CH$_4$ and partially as CO$_2$ at $\log f_{\mathrm{O_{2}}}<$IW$-$1 \citep{Li2015}. Future observations of exoplanetary atmospheric gases such as CO/CO$_2$ or CH$_4$ will not imply existence or absence of graphite-rich surfaces.

If the graphite layer is several hundreds of kilometers thick, it might not allow direct recycling of the mantle material to the surface. Such a graphite surface without essential life-bearing elements other than carbon will make the planet potentially uninhabitable. However, deep silicate volcanism, along with the presence of water, could still alter the surface composition of a C-enriched rocky exoplanet during the course of its evolution if penetration of material through the graphite is possible. To further assess these scenarios, detailed studies of the thermal and mechanical behaviour of graphite/diamond crusts are required. 

\section{Summary and conclusions}\label{paper2:conclusion}

We performed the first high-pressure high-temperature experiments on chemical mixtures representing bulk compositions of small C-enriched rocky exoplanets at 1 AU from their host star based on the calculations of a study modeling the chemistry in the protoplanetary disk of a high C/O star. Our results show that fully differentiated C-enriched rocky exoplanets consist of three major types of phases forming an iron-rich core, a silicate mantle and a graphite (and diamond) layer on top of the silicate mantle. Their mineralogy depends on oxygen fugacity and Mg/Si, Al/Si, Ca/Si, S/Fe and C/O ratios.

For S/Fe ratios in iron alloys between 0.1 and 0.8 and at pressures below $\sim4–6$~GPa, the core stratifies into a S-poor Fe inner core surrounded by a S-rich Fe outer core. The variety in mantle silicate minerals is largely independent of the C/O ratio. The sequential condensation model from \citep{Moriarty2014} at 1~AU from the host star result in C-enriched rocky exoplanets with higher oxygen fugacity conditions compared to the equilibrium condensation model. High C/O ratios in planet-forming refractory material do not necessarily imply reducing conditions as the amount of C has no direct impact on the oxygen fugacity. Extremely reducing (\textless IW$-$6) or oxidizing conditions (\textgreater IW$+$1) would be needed to stabilize silicon carbide or carbonates such as calcite and magnesite, respectively, in C-enriched planetary interiors. The minimum amount of carbon needed for carbon-saturation in the type of C-enriched rocky exoplanets considered in this study is 0.05$-$0.7~wt\% (molar C/O $\sim$ 0.002$-$0.03), which lies between the upper bounds of 200~ppm and 9~wt\% for mantle-only and core-only planets, respectively. 

Any amounts of carbon exceeding the carbon-saturation limit would be in the form of graphite. If the graphite layer is deep enough to exceed pressures of 2$-$15~GPa, depending on the temperature profile, a diamond layer would exist beneath the graphite layer. Carbon in the form of graphite can significantly affect the mass of an exoplanet for a fixed radius. For example, only a 10~wt\% graphite crust is sufficient to decrease the derived mass of Kepler-37b by 7\%, a difference detectable by future space missions focusing on determinations of both mass and radius of rocky exoplanets with insignificant gaseous envelopes. Rocky exoplanets with graphite-rich surfaces would appear dark in future observations because of low albedos due to graphite. Atmospheres of such planets are likely thin or non-existent, and the detection of CO/CO$_2$ or CH$_4$ on its own cannot confirm the presence or absence of a graphite-rich surface. Surfaces of such planets are less likely to be hospitable for life because of the lack of life-bearing elements other than carbon.

\section*{Acknowledgements}

We thank three anonymous reviewers for their constructive comments in improving this manuscript. This work is part of the Planetary and Exoplanetary Science Network (PEPSci), funded by the Netherlands Organization for Scientific Research (NWO, Project no. 648.001.005). We are grateful to Sergei Matveev and Tilly Bouten from Utrecht University for their technical assistance during EPMA measurements at Utrecht University. We thank Rajdeep Dasgupta for facilitating analyses of light elements in metals in the EPMA Laboratory at Rice University. We are also thankful to Jack Moriarty for providing data from their \citep{Moriarty2014} study.

\appendix

\section{Experimental sample assembly}\label{paper2:appendixSampleAssembly}

\begin{figure*}[!ht]
  \centering
  \medskip
  \includegraphics[width=.7\textwidth]{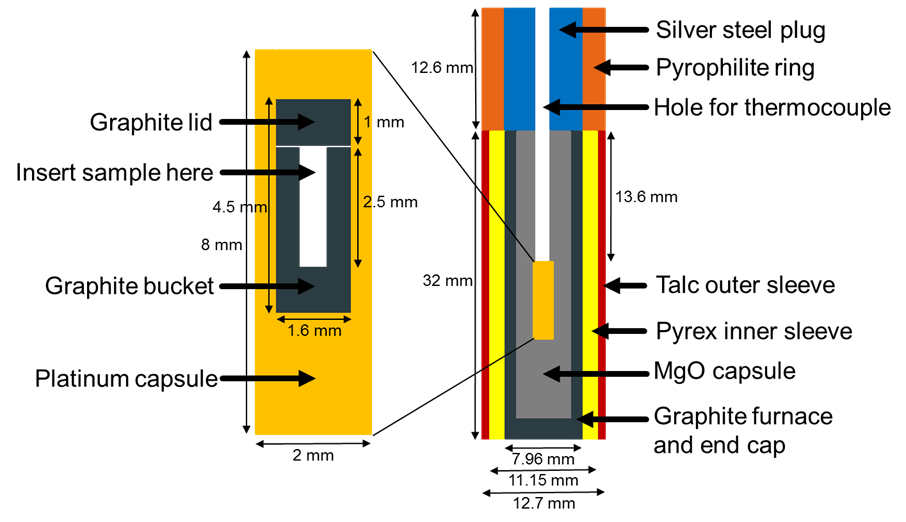}
  \caption[Sample Assembly]{Components of the sample assembly used to perform high-pressure experiments.}
  \label{fig:SampleAssembly}
\end{figure*}

Sample powder was inserted in a 1.6~mm-wide graphite capsule with a graphite lid (Fig.~\ref{fig:SampleAssembly}). This graphite capsule was put into a 2~mm-wide Pt capsule which was sealed and arc-welded on both ends using a Lampert PUK 3 welder. The Pt capsule was placed in a MgO rod sealed with MgO cement. The MgO rod was introduced in a graphite furnace, thermally insulated by surrounding it with an inner pyrex sleeve and an outer talc sleeve. A four-bore $\mathrm{Al_{2}O_{3}}$ rod through which thermocouple wires were threaded, was placed on the top of MgO rod. Pressure calibration of the assembly was performed by bracketing the albite to jadeite plus quartz and fayalite to ferrosilite plus quartz transitions \citep{VanKanParker2011}. The resulting pressure correction of 3\% is consistent with literature data \citep{McDade2002}. On top of the talc-pyrex assembly, a hardened silver steel plug with a pyrophillite ring and a hole for thermocouple was placed. A $\mathrm{W_{97}Re_{3}/W_{75}Re_{25}}$ (type D) thermocouple was placed in the thermocouple hole directly above the Pt capsule. The distance of 1$-$3.5~mm between the thermocouple tip and the sample produced a temperature difference of 10~K \citep{Watson2002}.

\section{Oxygen fugacity calculations}\label{paper2:appendixOxygenFugacity}

We computed oxygen fugacity ($f_{\mathrm{O_{2}}}$) in our experiments with respect to the iron-w{\"u}stite (IW) buffer by using the following equation: 

\begin{equation}\label{eq:fO2} 
\log f_{\mathrm{O_{2}}} (\Delta \mathrm{IW}) = 2 \log \frac{X^{\mathrm{sil}}_{\mathrm{FeO}} \cdot \gamma^{\mathrm{sil}}_{\mathrm{FeO}}}{X^{\mathrm{alloy}}_{\mathrm{Fe}} \cdot \gamma^{\mathrm{alloy}}_{\mathrm{Fe}}},
\end{equation} where $X^{\mathrm{sil}}_{\mathrm{FeO}}$ and $\gamma^{\mathrm{sil}}_{\mathrm{FeO}}$ are the mole fraction and the activity of FeO in silicate melt, and $X^{\mathrm{alloy}}_{\mathrm{Fe}}$ and $\gamma^{\mathrm{alloy}}_{\mathrm{Fe}}$ are the mole fraction and the activity of Fe in S-rich Fe alloy. We assumed a non-ideal solution behavior of silicate melt and iron alloy, which implies non-unity values for $\gamma^{\mathrm{sil}}_{\mathrm{FeO}}$ and $\gamma^{\mathrm{alloy}}_{\mathrm{Fe}}$. A fixed value of $\gamma^{\mathrm{sil}}_{\mathrm{FeO}}=1.5$, the average from the two studies that determined
$\gamma^{\mathrm{sil}}_{\mathrm{FeO}}$ for a wide range of melt compositions were used
\citep{Holzheid1997,ONeill2002}, assuming no significant pressure effect on $\gamma^{\mathrm{sil}}_{\mathrm{FeO}}$ \citep{Toplis2005}. For $\gamma^{\mathrm{alloy}}_{\mathrm{Fe}}$, we computed $\gamma^{\mathrm{alloy}}_{\mathrm{Fe}}$ from \citet{Lee2002} using 

\begin{eqnarray}\label{eq:gamFe}
\left. \ln \gamma^{\mathrm{alloy}}_{\mathrm{Fe}} = \frac{\alpha_{2}}{2} (1-X^{\mathrm{alloy}}_{\mathrm{Fe}})^{2} \right. \\ \nonumber
+ \left. \frac{\alpha_{3}}{3} (1-X^{\mathrm{alloy}}_{\mathrm{Fe}})^{3} + \frac{\alpha_{4}}{4} (1-X^{\mathrm{alloy}}_{\mathrm{Fe}})^{4}, \right.
\end{eqnarray} where $\alpha_{2}=3.80$, $\alpha_{3}=-5.24$ and $\alpha_{4}=2.58$ at 1623~K, $\alpha_{2}=4.01$, $\alpha_{3}=-5.52$ and $\alpha_{4}=2.71$ at 1723~K and $\alpha_{2}=4.25$, $\alpha_{3}=-5.84$ and $\alpha_{4}=2.87$ at 1823~K. In the absence of silicate melts, as in some of our experiments, we used $X_{\mathrm{FeO}}$ of olivine instead. Our oxygen fugacity calculations are given in Table \ref{tab:ExperimentConditions}.

\section{Mineral-melt equilibrium}\label{paper2:appendixEquilibrium}

To assess mineral-melt equilibrium, we calculated olivine-melt and orthopyroxene-melt Fe-Mg exchange coefficients, $K_{D}$ and $K'_{D}$, following \citet{Kushiro1998} and \citet{Toplis2005}, respectively. 

\begin{eqnarray}\label{eq:KD} 
K_{D} \ ^{\mathrm{Mg-Fe}}_{\mathrm{Olv-Melt}} = \frac{X^{\mathrm{Mg}}_{\mathrm{Melt}}/X^{\mathrm{Mg}}_{\mathrm{Olv}}}{X^{\mathrm{Fe}}_{\mathrm{Melt}}/X^{\mathrm{Fe}}_{\mathrm{Olv}}}
\end{eqnarray} 

\begin{eqnarray}\label{eq:K1D} 
K'_{D} \ ^{\mathrm{Mg-Fe}}_{\mathrm{Olv-Melt}} = \mathrm{exp} \left[ \left( \frac{-6766}{RT} - \frac{7.34}{R} \right) \right.  \\ \nonumber
+ \left. \mathrm{ln}[0.036 X^{\mathrm{SiO_{2}}}_{\mathrm{Melt}} - 0.22] \right.  \\ \nonumber
+ \left. \left( \frac{3000 (1 - 2 Y^{\mathrm{Mg/(Mg+Fe)}}_{\mathrm{Olv}})}{RT} \right) \right. \\ \nonumber
+ \left. \left( \frac{0.035 (P - 1)}{RT} \right) \right] 
\end{eqnarray} where $X^{\mathrm{a}}_{\mathrm{b}}$ is mol\% of a in b, $Y^{\mathrm{Mg/(Mg+Fe)}}_{\mathrm{Olv}}$ is molar Mg/(Mg+Fe) in olivine, $R$ is the gas constant, $T$ is temperature in K and $P$ is pressure in bar. Our calculations of $K_{D}$ and $K'_{D}$ for all runs result in values between 0.22$-$0.38 and 0.24$-$0.28, respectively (see Table \ref{tab:PhaseCompositionSilicates}). According to \citet{Toplis2005}, this range is consistent with equilibrium. For orthopyroxene in the TC run at 2~GPa and 1823~K, and the EC runs at 1~GPa and 1723~K and 2~GPa and 1823~K, $K_{D}$ ranges between 0.21$-$0.32, also within the acceptable range for equilibrated systems. 


%\bibliographystyle{aa}
%\bibliographystyle{aas_macros}
\bibliographystyle{icarus}
\bibliography{CarbonExoplanets}

\begin{thebibliography}{}

\bibitem[{Anderson} et~al.(2017){Anderson}, {Bergin}, {Blake}, {Ciesla},
  {Visser}, and {Lee}]{Anderson2017}
{Anderson}, D.~E., {Bergin}, E.~A., {Blake}, G.~A., {Ciesla}, F.~J., {Visser},
  R., {Lee}, J.-E., 2017.
\newblock {Destruction of refractory carbon in protoplanetary disks}.
\newblock \apj~845, 13.

\bibitem[{Armstrong}(1995){Armstrong}]{Armstrong1995}
{Armstrong}, J.~T., 1995.
\newblock {CITZAF: A package of correction programs for the quantitative
  electron microbeam X-ray analysis of thick polished materials, thin films,
  and particles}.
\newblock Microbeam Analysis~4 (3), 177--200.

\bibitem[{Batalha}(2014){Batalha}]{Batalha2014}
{Batalha}, N.~M., 2014.
\newblock {Exploring exoplanet populations with NASA's Kepler Mission}.
\newblock Proceedings of the National Academy of Science~111, 12647--12654.

\bibitem[{Birch}(1947){Birch}]{Birch1947}
{Birch}, F., 1947.
\newblock {Finite elastic strain of cubic crystals}.
\newblock Physical Review~71, 809--824.

\bibitem[{Bizzarro} et~al.(2005){Bizzarro}, {Baker}, {Haack}, and
  {Lundgaard}]{Bizzarro2005}
{Bizzarro}, M., {Baker}, J.~A., {Haack}, H., {Lundgaard}, K.~L., 2005.
\newblock {Rapid timescales for accretion and melting of differentiated
  planetesimals inferred from $^{26}$Al-$^{26}$Mg chronometry}.
\newblock \apjl~632, L41--L44.

\bibitem[{Bond} et~al.(2010b){Bond}, {O'Brien}, and {Lauretta}]{Bond2010b}
{Bond}, J.~C., {O'Brien}, D.~P., {Lauretta}, D.~S., 2010b.
\newblock {The compositional diversity of extrasolar terrestrial planets. I. In
  situ simulations}.
\newblock \apj~715, 1050--1070.

\bibitem[{Boujibar} et~al.(2014){Boujibar}, {Andrault}, {Bouhifd},
  {Bolfan-Casanova}, {Devidal}, and {Trcera}]{Boujibar2014}
{Boujibar}, A., {Andrault}, D., {Bouhifd}, M.~A., {Bolfan-Casanova}, N.,
  {Devidal}, J.-L., {Trcera}, N., 2014.
\newblock {Metal-silicate partitioning of sulphur, new experimental and
  thermodynamic constraints on planetary accretion}.
\newblock Earth and Planetary Science Letters~391, 42--54.

\bibitem[{Brewer} et~al.(2016){Brewer}, {Fischer}, {Valenti}, and
  {Piskunov}]{Brewer2016}
{Brewer}, J.~M., {Fischer}, D.~A., {Valenti}, J.~A., {Piskunov}, N., 2016.
\newblock {Spectral properties of cool stars: Extended abundance analysis of
  1,617 planet-search stars}.
\newblock \apjs~225, 32.

\bibitem[{Carter-Bond} et~al.(2012a){Carter-Bond}, {O'Brien}, {Delgado Mena},
  {Israelian}, {Santos}, and {Gonz{\'a}lez Hern{\'a}ndez}]{CarterBond2012a}
{Carter-Bond}, J.~C., {O'Brien}, D.~P., {Delgado Mena}, E., {Israelian}, G.,
  {Santos}, N.~C., {Gonz{\'a}lez Hern{\'a}ndez}, J.~I., 2012a.
\newblock {Low Mg/Si planetary host stars and their Mg-depleted terrestrial
  planets}.
\newblock \apjl~747, L2.

\bibitem[{Carter-Bond} et~al.(2012b){Carter-Bond}, {O'Brien}, and
  {Raymond}]{CarterBond2012b}
{Carter-Bond}, J.~C., {O'Brien}, D.~P., {Raymond}, S.~N., 2012b.
\newblock {The compositional diversity of extrasolar terrestrial planets. II.
  Migration simulations}.
\newblock \apj~760, 44.

\bibitem[{Chi} et~al.(2014){Chi}, {Dasgupta}, {Duncan}, and {Shimizu}]{Chi2014}
{Chi}, H., {Dasgupta}, R., {Duncan}, M.~S., {Shimizu}, N., 2014.
\newblock {Partitioning of carbon between Fe-rich alloy melt and silicate melt
  in a magma ocean - Implications for the abundance and origin of volatiles in
  Earth, Mars, and the Moon}.
\newblock \gca~139, 447--471.

\bibitem[{Colonna} et~al.(2011){Colonna}, {Fasolino}, and
  {Meijer}]{Colonna2011}
{Colonna}, F., {Fasolino}, A., {Meijer}, E.~J., 2011.
\newblock High-pressure high-temperature equation of state of graphite from
  monte carlo simulations.
\newblock Carbon~49, 364--368.

\bibitem[{Corgne} et~al.(2008){Corgne}, {Wood}, and {Fei}]{Corgne2008}
{Corgne}, A., {Wood}, B.~J., {Fei}, Y., 2008.
\newblock {C- and S-rich molten alloy immiscibility and core formation of
  planetesimals}.
\newblock \gca~72, 2409--2416.

\bibitem[{Dasgupta}(2013){Dasgupta}]{Dasgupta2013a}
{Dasgupta}, R., 2013.
\newblock {Ingassing, storage, and outgassing of terrestrial carbon through
  geologic time}.
\newblock Reviews in Mineralogy and Geochemistry~75, 183--229.

\bibitem[{Dasgupta} et~al.(2009){Dasgupta}, {Buono}, {Whelan}, and
  {Walker}]{Dasgupta2009}
{Dasgupta}, R., {Buono}, A., {Whelan}, G., {Walker}, D., 2009.
\newblock {High-pressure melting relations in Fe-C-S systems: Implications for
  formation, evolution, and structure of metallic cores in planetary bodies}.
\newblock \gca~73, 6678--6691.

\bibitem[{Dasgupta} et~al.(2013){Dasgupta}, {Chi}, {Shimizu}, {Buono}, and
  {Walker}]{Dasgupta2013b}
{Dasgupta}, R., {Chi}, H., {Shimizu}, N., {Buono}, A.~S., {Walker}, D., 2013.
\newblock {Carbon solution and partitioning between metallic and silicate melts
  in a shallow magma ocean: Implications for the origin and distribution of
  terrestrial carbon}.
\newblock \gca~102, 191--212.

\bibitem[{Dasgupta} and {Walker}(2008){Dasgupta} and {Walker}]{Dasgupta2008}
{Dasgupta}, R., {Walker}, D., 2008.
\newblock {Carbon solubility in core melts in a shallow magma ocean environment
  and distribution of carbon between the Earth's core and the mantle}.
\newblock \gca~72, 4627--4641.

\bibitem[{Delgado Mena} et~al.(2010){Delgado Mena}, {Israelian}, {Gonz{\'a}lez
  Hern{\'a}ndez}, {Bond}, {Santos}, {Udry}, and {Mayor}]{DelgadoMena2010}
{Delgado Mena}, E., et~al., 2010.
\newblock {Chemical clues on the formation of planetary systems: C/O Versus
  Mg/Si for HARPS GTO Sample}.
\newblock \apj~725, 2349--2358.

\bibitem[{Deng} et~al.(2013){Deng}, {Fei}, {Liu}, {Gong}, and
  {Shahar}]{Deng2013}
{Deng}, L., {Fei}, Y., {Liu}, X., {Gong}, Z., {Shahar}, A., 2013.
\newblock {Effect of carbon, sulfur and silicon on iron melting at high
  pressure: Implications for composition and evolution of the planetary
  terrestrial cores}.
\newblock \gca~114, 220--233.

\bibitem[{Dorn} et~al.(2015){Dorn}, {Khan}, {Heng}, {Connolly}, {Alibert},
  {Benz}, and {Tackley}]{Dorn2015}
{Dorn}, C., et~al., 2015.
\newblock {Can we constrain the interior structure of rocky exoplanets from
  mass and radius measurements?}
\newblock \aap~577, A83.

\bibitem[{Duncan} et~al.(2017){Duncan}, {Dasgupta}, and {Tsuno}]{Duncan2017}
{Duncan}, M.~S., {Dasgupta}, R., {Tsuno}, K., 2017.
\newblock {Experimental determination of CO$_{2}$ content at graphite
  saturation along a natural basalt-peridotite melt join: Implications for the
  fate of carbon in terrestrial magma oceans}.
\newblock Earth and Planetary Science Letters~466, 115--128.

\bibitem[{Elkins-Tanton}(2012){Elkins-Tanton}]{ElkinsTanton2012}
{Elkins-Tanton}, L.~T., 2012.
\newblock {Magma oceans in the inner solar system}.
\newblock Annual Review of Earth and Planetary Sciences~40, 113--139.

\bibitem[{Fei} et~al.(2016){Fei}, {Murphy}, {Shibazaki}, {Shahar}, and
  {Huang}]{Fei2016}
{Fei}, Y., {Murphy}, C., {Shibazaki}, Y., {Shahar}, A., {Huang}, H., 2016.
\newblock {Thermal equation of state of hcp-iron: Constraint on the density
  deficit of Earth's solid inner core}.
\newblock \grl~43, 6837--6843.

\bibitem[{Fortier} et~al.(2014){Fortier}, {Beck}, {Benz}, {Broeg}, {Cessa},
  {Ehrenreich}, and {Thomas}]{Fortier2014}
{Fortier}, A., et~al., 2014.
\newblock {CHEOPS: a space telescope for ultra-high precision photometry of
  exoplanet transits}.
\newblock In: Space Telescopes and Instrumentation 2014: Optical, Infrared, and
  Millimeter Wave, Volume 9143 of {\em \procspie}, pp.\  91432J.

\bibitem[{Ghiringhelli} et~al.(2005){Ghiringhelli}, {Los}, {Meijer},
  {Fasolino}, and {Frenkel}]{Ghiringhelli2005}
{Ghiringhelli}, L.~M., {Los}, J.~H., {Meijer}, E.~J., {Fasolino}, A.,
  {Frenkel}, D., 2005.
\newblock {Modeling the phase diagram of carbon}.
\newblock Physical Review Letters~94 (14), 145701.

\bibitem[{Giordano} et~al.(2008){Giordano}, {Russell}, and
  {Dingwell}]{Giordano2008}
{Giordano}, D., {Russell}, J.~K., {Dingwell}, D.~B., 2008.
\newblock {Viscosity of magmatic liquids: A model}.
\newblock Earth and Planetary Science Letters~271, 123--134.

\bibitem[{Hakim} et~al.(2018a){Hakim}, {Rivoldini}, {Van Hoolst}, {Cottenier},
  {Jaeken}, {Chust}, and {Steinle-Neumann}]{Hakim2018a}
{Hakim}, K., et~al., 2018a.
\newblock {A new ab initio equation of state of hcp-Fe and its implication on
  the interior structure and mass-radius relations of rocky super-Earths}.
\newblock \icarus~313, 61--78.

\bibitem[{Hakim} et~al.(2018b){Hakim}, {van Westrenen}, and
  {Dominik}]{Hakim2018b}
{Hakim}, K., {van Westrenen}, W., {Dominik}, C., 2018b.
\newblock {Capturing the oxidation of silicon carbide in rocky exoplanetary
  interiors}.
\newblock \aap~618, L6.

\bibitem[{Hashizume} and {Sugiura}(1998){Hashizume} and
  {Sugiura}]{Hashizume1998}
{Hashizume}, K., {Sugiura}, N., 1998.
\newblock {Transportation of gaseous elements and isotopes in a thermally
  evolving chondritic planetesimal}.
\newblock Meteoritics and Planetary Science~33, 1181--1195.

\bibitem[{Hevey} and {Sanders}(2006){Hevey} and {Sanders}]{Hevey2006}
{Hevey}, P.~J., {Sanders}, I.~S., 2006.
\newblock {A model for planetesimal meltdown by $^{26}$Al and its implications
  for meteorite parent bodies}.
\newblock Meteoritics and Planetary Science~41, 95--106.

\bibitem[{Hirose} and {Fei}(2002){Hirose} and {Fei}]{HiroseFei2002}
{Hirose}, K., {Fei}, Y., 2002.
\newblock {Subsolidus and melting phase relations of basaltic composition in
  the uppermostlower mantle}.
\newblock \gca~66, 2099--2108.

\bibitem[{Holzheid} et~al.(1997){Holzheid}, {Palme}, and
  {Chakraborty}]{Holzheid1997}
{Holzheid}, A., {Palme}, H., {Chakraborty}, S., 1997.
\newblock {The activities of NiO, CoO and FeO in silicate melts}.
\newblock Chemical Geology~139, 21--38.

\bibitem[{Javoy} et~al.(2010){Javoy}, {Kaminski}, {Guyot}, {Andrault},
  {Sanloup}, {Moreira}, {Labrosse}, {Jambon}, {Agrinier}, {Davaille}, and
  {Jaupart}]{Javoy2010}
{Javoy}, M., et~al., 2010.
\newblock {The chemical composition of the Earth: Enstatite chondrite models}.
\newblock Earth and Planetary Science Letters~293, 259--268.

\bibitem[{Johansen} et~al.(2007){Johansen}, {Oishi}, {Mac Low}, {Klahr},
  {Henning}, and {Youdin}]{Johansen2007}
{Johansen}, A., {Oishi}, J.~S., {Mac Low}, M.-M., {Klahr}, H., {Henning}, T.,
  {Youdin}, A., 2007.
\newblock {Rapid planetesimal formation in turbulent circumstellar disks}.
\newblock \nat~448, 1022--1025.

\bibitem[{Jugo} et~al.(2005){Jugo}, {Luth}, and {Richards}]{Jugo2005}
{Jugo}, P.~J., {Luth}, R.~W., {Richards}, J.~P., 2005.
\newblock {Experimental data on the speciation of sulfur as a function of
  oxygen fugacity in basaltic melts}.
\newblock \gca~69, 497--503.

\bibitem[{Jugo} et~al.(2010){Jugo}, {Wilke}, and {Botcharnikov}]{Jugo2010}
{Jugo}, P.~J., {Wilke}, M., {Botcharnikov}, R.~E., 2010.
\newblock {Sulfur K-edge XANES analysis of natural and synthetic basaltic
  glasses: Implications for S speciation and S content as function of oxygen
  fugacity}.
\newblock \gca~74, 5926--5938.

\bibitem[{Klarmann} et~al.(2018){Klarmann}, {Ormel}, and
  {Dominik}]{Klarmann2018}
{Klarmann}, L., {Ormel}, C.~W., {Dominik}, C., 2018.
\newblock {Radial and vertical dust transport inhibit refractory carbon
  depletion in protoplanetary disks}.
\newblock \aap~618, L1.

\bibitem[{Kruijer} et~al.(2013){Kruijer}, {Fischer-G{\"o}dde}, {Kleine},
  {Sprung}, {Leya}, and {Wieler}]{Kruijer2013}
{Kruijer}, T.~S., {Fischer-G{\"o}dde}, M., {Kleine}, T., {Sprung}, P., {Leya},
  I., {Wieler}, R., 2013.
\newblock {Neutron capture on Pt isotopes in iron meteorites and the Hf-W
  chronology of core formation in planetesimals}.
\newblock Earth and Planetary Science Letters~361, 162--172.

\bibitem[{Kuchner} and {Seager}(2005){Kuchner} and {Seager}]{Kuchner2005}
{Kuchner}, M.~J., {Seager}, S., 2005.
\newblock {Extrasolar carbon planets}.
\newblock ArXiv Astrophysics e-prints.

\bibitem[{Kushiro} and {Walter}(1998){Kushiro} and {Walter}]{Kushiro1998}
{Kushiro}, I., {Walter}, M.~J., 1998.
\newblock {Mg-Fe partitioning between olivine and mafic-ultramafic melts}.
\newblock \grl~25, 2337--2340.

\bibitem[{Lazar} et~al.(2014){Lazar}, {Zhang}, {Manning}, and
  {Mysen}]{Lazar2014}
{Lazar}, C., {Zhang}, C., {Manning}, C.~E., {Mysen}, B.~O., 2014.
\newblock {Redox effects on calcite-portlandite-fluid equilibria at forearc
  conditions: Carbon mobility, methanogenesis, and reduction melting of
  calcite}.
\newblock American Mineralogist~99, 1604--1615.

\bibitem[{Lee} and {Morita}(2002){Lee} and {Morita}]{Lee2002}
{Lee}, J., {Morita}, K., 2002.
\newblock Evaluation of surface tension and adsorption for liquid fe-s alloys.
\newblock ISIJ International~42 (6), 588--594.

\bibitem[{Lee} et~al.(2010){Lee}, {Bergin}, and {Nomura}]{Lee2010}
{Lee}, J.-E., {Bergin}, E.~A., {Nomura}, H., 2010.
\newblock {The solar nebula on fire: A solution to the carbon deficit in the
  inner solar system}.
\newblock \apjl~710, L21--L25.

\bibitem[{L{\'e}ger} et~al.(2009){L{\'e}ger}, {Rouan}, {Schneider}, {Barge},
  {Fridlund}, {Samuel}, {Ollivier}, {Guenther}, {Deleuil}, {Deeg}, {Auvergne},
  {Alonso}, {Aigrain}, {Alapini}, {Almenara}, {Baglin}, {Barbieri}, {Bruntt},
  {Bord{\'e}}, {Bouchy}, {Cabrera}, {Catala}, {Carone}, {Carpano}, {Csizmadia},
  {Dvorak}, {Erikson}, {Ferraz-Mello}, {Foing}, {Fressin}, {Gandolfi},
  {Gillon}, {Gondoin}, {Grasset}, {Guillot}, {Hatzes}, {H{\'e}brard}, {Jorda},
  {Lammer}, {Llebaria}, {Loeillet}, {Mayor}, {Mazeh}, {Moutou}, {P{\"a}tzold},
  {Pont}, {Queloz}, {Rauer}, {Renner}, {Samadi}, {Shporer}, {Sotin}, {Tingley},
  {Wuchterl}, {Adda}, {Agogu}, {Appourchaux}, {Ballans}, {Baron}, {Beaufort},
  {Bellenger}, {Berlin}, {Bernardi}, {Blouin}, {Baudin}, {Bodin}, {Boisnard},
  {Boit}, {Bonneau}, {Borzeix}, {Briet}, {Buey}, {Butler}, {Cailleau},
  {Cautain}, {Chabaud}, {Chaintreuil}, {Chiavassa}, {Costes}, {Cuna Parrho},
  {de Oliveira Fialho}, {Decaudin}, {Defise}, {Djalal}, {Epstein}, {Exil},
  {Faur{\'e}}, {Fenouillet}, {Gaboriaud}, {Gallic}, {Gamet}, {Gavalda},
  {Grolleau}, {Gruneisen}, {Gueguen}, {Guis}, {Guivarc'h}, {Guterman},
  {Hallouard}, {Hasiba}, {Heuripeau}, {Huntzinger}, {Hustaix}, {Imad},
  {Imbert}, {Johlander}, {Jouret}, {Journoud}, {Karioty}, {Kerjean},
  {Lafaille}, {Lafond}, {Lam-Trong}, {Landiech}, {Lapeyrere}, {Larqu{\'e}},
  {Laudet}, {Lautier}, {Lecann}, {Lefevre}, {Leruyet}, {Levacher}, {Magnan},
  {Mazy}, {Mertens}, {Mesnager}, {Meunier}, {Michel}, {Monjoin}, {Naudet},
  {Nguyen-Kim}, {Orcesi}, {Ottacher}, {Perez}, {Peter}, {Plasson}, {Plesseria},
  {Pontet}, {Pradines}, {Quentin}, {Reynaud}, {Rolland}, {Rollenhagen},
  {Romagnan}, {Russ}, {Schmidt}, {Schwartz}, {Sebbag}, {Sedes}, {Smit},
  {Steller}, {Sunter}, {Surace}, {Tello}, {Tiph{\`e}ne}, {Toulouse}, {Ulmer},
  {Vandermarcq}, {Vergnault}, {Vuillemin}, and {Zanatta}]{Leger2009}
{L{\'e}ger}, A., et~al., 2009.
\newblock {Transiting exoplanets from the CoRoT space mission. VIII. CoRoT-7b:
  the first super-Earth with measured radius}.
\newblock \aap~506, 287--302.

\bibitem[{Li} et~al.(2015){Li}, {Dasgupta}, and {Tsuno}]{Li2015}
{Li}, Y., {Dasgupta}, R., {Tsuno}, K., 2015.
\newblock {The effects of sulfur, silicon, water, and oxygen fugacity on carbon
  solubility and partitioning in Fe-rich alloy and silicate melt systems at 3
  GPa and 1600 {$\deg$}C: Implications for core-mantle differentiation and
  degassing of magma oceans and reduced planetary mantles}.
\newblock Earth and Planetary Science Letters~415, 54--66.

\bibitem[{Li} et~al.(2016){Li}, {Dasgupta}, {Tsuno}, {Monteleone}, and
  {Shimizu}]{Li2016}
{Li}, Y., {Dasgupta}, R., {Tsuno}, K., {Monteleone}, B., {Shimizu}, N., 2016.
\newblock {Carbon and sulfur budget of the silicate Earth explained by
  accretion of differentiated planetary embryos}.
\newblock Nature Geoscience~9, 781--785.

\bibitem[{Lord} et~al.(2009){Lord}, {Walter}, {Dasgupta}, {Walker}, and
  {Clark}]{Lord2009}
{Lord}, O.~T., {Walter}, M.~J., {Dasgupta}, R., {Walker}, D., {Clark}, S.~M.,
  2009.
\newblock {Melting in the Fe-C system to 70 GPa}.
\newblock Earth and Planetary Science Letters~284, 157--167.

\bibitem[{Madhusudhan} et~al.(2012){Madhusudhan}, {Lee}, and
  {Mousis}]{Madhusudhan2012}
{Madhusudhan}, N., {Lee}, K.~K.~M., {Mousis}, O., 2012.
\newblock {A possible carbon-rich interior in super-Earth 55 Cancri e}.
\newblock \apjl~759, L40.

\bibitem[{Marty} et~al.(2013){Marty}, {Alexander}, and {Raymond}]{Marty2013}
{Marty}, B., {Alexander}, C.~M.~O., {Raymond}, S.~N., 2013.
\newblock {Primordial origins of Earth's carbon}.
\newblock Reviews in Mineralogy and Geochemistry~75, 149--181.

\bibitem[{McDade} et~al.(2002){McDade}, {Wood}, {Van Westrenen}, {Brooker},
  {Gudmundsson}, {Soulard}, {Najorka}, and {Blundy}]{McDade2002}
{McDade}, P., et~al., 2002.
\newblock {Pressure corrections for a selection of piston-cylinder cell
  assemblies}.
\newblock Mineralogical Magazine~66, 1021--1028.

\bibitem[{Morard} and {Katsura}(2010){Morard} and {Katsura}]{Morard2010}
{Morard}, G., {Katsura}, T., 2010.
\newblock {Pressure-temperature cartography of Fe-S-Si immiscible system}.
\newblock \gca~74, 3659--3667.

\bibitem[{Moriarty} et~al.(2014){Moriarty}, {Madhusudhan}, and
  {Fischer}]{Moriarty2014}
{Moriarty}, J., {Madhusudhan}, N., {Fischer}, D., 2014.
\newblock {Chemistry in an evolving protoplanetary disk: Effects on terrestrial
  planet composition}.
\newblock \apj~787, 81.

\bibitem[{Nabiei} et~al.(2018){Nabiei}, {Badro}, {Dennenwaldt}, {Oveisi},
  {Cantoni}, {H{\'e}bert}, {El Goresy}, {Barrat}, and {Gillet}]{Nabiei2018}
{Nabiei}, F., et~al., 2018.
\newblock {A large planetary body inferred from diamond inclusions in a
  ureilite meteorite}.
\newblock Nature Communications~9, 1327.

\bibitem[{Nakajima} and {Sorahana}(2016){Nakajima} and
  {Sorahana}]{Nakajima2016}
{Nakajima}, T., {Sorahana}, S., 2016.
\newblock {Carbon-to-oxygen Ratios in M Dwarfs and Solar-type Stars}.
\newblock \apj~830, 159.

\bibitem[{Nisr} et~al.(2017){Nisr}, {Meng}, {MacDowell}, {Yan}, {Prakapenka},
  and {Shim}]{Nisr2017}
{Nisr}, C., {Meng}, Y., {MacDowell}, A.~A., {Yan}, J., {Prakapenka}, V.,
  {Shim}, S.-H., 2017.
\newblock {Thermal expansion of SiC at high pressure-temperature and
  implications for thermal convection in the deep interiors of carbide
  exoplanets}.
\newblock Journal of Geophysical Research (Planets)~122, 124--133.

\bibitem[{O'Neill} and {Eggins}(2002){O'Neill} and {Eggins}]{ONeill2002}
{O'Neill}, H.~S.~C., {Eggins}, S.~M., 2002.
\newblock {The effect of melt composition on trace element partitioning: an
  experimental investigation of the activity coefficients of FeO, NiO, CoO,
  MoO2 and MoO3 in silicate melts}.
\newblock Chemical Geology~186, 151--181.

\bibitem[{Peplowski} et~al.(2016){Peplowski}, {Klima}, {Lawrence}, {Ernst},
  {Denevi}, {Frank}, {Goldsten}, {Murchie}, {Nittler}, and
  {Solomon}]{Peplowski2016}
{Peplowski}, P.~N., et~al., 2016.
\newblock {Remote sensing evidence for an ancient carbon-bearing crust on
  Mercury}.
\newblock Nature Geoscience~9, 273--276.

\bibitem[{Petigura} and {Marcy}(2011){Petigura} and {Marcy}]{Petigura2011}
{Petigura}, E.~A., {Marcy}, G.~W., 2011.
\newblock {Carbon and oxygen in nearby stars: Keys to protoplanetary disk
  chemistry}.
\newblock \apj~735, 41.

\bibitem[{Ragazzoni} et~al.(2016){Ragazzoni}, {Magrin}, {Rauer}, {Pagano},
  {Nascimbeni}, {Piotto}, {Piazza}, {Levacher}, {Schweitzer}, {Basso}, {Bandy},
  {Benz}, {Bergomi}, {Biondi}, {Boerner}, {Borsa}, {Brandeker}, {Br{\"a}ndli},
  {Bruno}, {Cabrera}, {Chinellato}, {De Roche}, {Dima}, {Erikson}, {Farinato},
  {Munari}, {Ghigo}, {Greggio}, {Gullieuszik}, {Klebor}, {Marafatto},
  {Mogulsky}, {Peter}, {Rieder}, {Sicilia}, {Spiga}, {Viotto}, {Wieser},
  {Heras}, {Gondoin}, {Bodin}, and {Catala}]{Ragazzoni2016}
{Ragazzoni}, R., et~al., 2016.
\newblock {PLATO: a multiple telescope spacecraft for exo-planets hunting}.
\newblock In: Space Telescopes and Instrumentation 2016: Optical, Infrared, and
  Millimeter Wave, Volume 9904 of {\em \procspie}, pp.\  990428.

\bibitem[{Rai} and {Westrenen}(2013){Rai} and {Westrenen}]{Rai2013}
{Rai}, N., {Westrenen}, W., 2013.
\newblock {Core-mantle differentiation in Mars}.
\newblock Journal of Geophysical Research (Planets)~118, 1195--1203.

\bibitem[{Ricker} et~al.(2014){Ricker}, {Winn}, {Vanderspek}, {Latham},
  {Bakos}, {Bean}, {Berta-Thompson}, {Brown}, {Buchhave}, {Butler}, {Butler},
  {Chaplin}, {Charbonneau}, {Christensen-Dalsgaard}, {Clampin}, {Deming},
  {Doty}, {De Lee}, {Dressing}, {Dunham}, {Endl}, {Fressin}, {Ge}, {Henning},
  {Holman}, {Howard}, {Ida}, {Jenkins}, {Jernigan}, {Johnson}, {Kaltenegger},
  {Kawai}, {Kjeldsen}, {Laughlin}, {Levine}, {Lin}, {Lissauer}, {MacQueen},
  {Marcy}, {McCullough}, {Morton}, {Narita}, {Paegert}, {Palle}, {Pepe},
  {Pepper}, {Quirrenbach}, {Rinehart}, {Sasselov}, {Sato}, {Seager},
  {Sozzetti}, {Stassun}, {Sullivan}, {Szentgyorgyi}, {Torres}, {Udry}, and
  {Villasenor}]{Ricker2014}
{Ricker}, G.~R., et~al., 2014.
\newblock {Transiting Exoplanet Survey Satellite (TESS)}.
\newblock In: Space telescopes and instrumentation 2014: Optical, infrared, and
  millimeter wave, Volume 9143 of {\em \procspie}, pp.\  914320.

\bibitem[{Rohrbach} et~al.(2014){Rohrbach}, {Ghosh}, {Schmidt}, {Wijbrans}, and
  {Klemme}]{Rohrbach2014}
{Rohrbach}, A., {Ghosh}, S., {Schmidt}, M.~W., {Wijbrans}, C.~H., {Klemme}, S.,
  2014.
\newblock {The stability of Fe-Ni carbides in the Earth{\#700}s mantle:
  Evidence for a low Fe-Ni-C melt fraction in the deep mantle}.
\newblock Earth and Planetary Science Letters~388, 211--221.

\bibitem[{Rohrbach} and {Schmidt}(2011){Rohrbach} and {Schmidt}]{Rohrbach2011}
{Rohrbach}, A., {Schmidt}, M.~W., 2011.
\newblock {Redox freezing and melting in the Earth's deep mantle resulting from
  carbon-iron redox coupling}.
\newblock \nat~472, 209--212.

\bibitem[{Santos} et~al.(2017){Santos}, {Adibekyan}, {Dorn}, {Mordasini},
  {Noack}, {Barros}, {Delgado-Mena}, {Demangeon}, {Faria}, {Israelian}, and
  {Sousa}]{Santos2017}
{Santos}, N.~C., et~al., 2017.
\newblock {Constraining planet structure and composition from stellar
  chemistry: trends in different stellar populations}.
\newblock \aap~608, A94.

\bibitem[{Sata} et~al.(2010){Sata}, {Hirose}, {Shen}, {Nakajima}, {Ohishi}, and
  {Hirao}]{Sata2010}
{Sata}, N., {Hirose}, K., {Shen}, G., {Nakajima}, Y., {Ohishi}, Y., {Hirao},
  N., 2010.
\newblock {Compression of FeSi, Fe$_{3}$C, Fe$_{0.95}$O, and FeS under the core
  pressures and implication for light element in the Earth's core}.
\newblock Journal of Geophysical Research (Solid Earth)~115, B09204.

\bibitem[{Sch{\"a}fer} et~al.(2017){Sch{\"a}fer}, {Yang}, and
  {Johansen}]{SchaferJohansen2017}
{Sch{\"a}fer}, U., {Yang}, C.-C., {Johansen}, A., 2017.
\newblock {Initial mass function of planetesimals formed by the streaming
  instability}.
\newblock \aap~597, A69.

\bibitem[{Seager} et~al.(2007){Seager}, {Kuchner}, {Hier-Majumder}, and
  {Militzer}]{Seager2007}
{Seager}, S., {Kuchner}, M., {Hier-Majumder}, C.~A., {Militzer}, B., 2007.
\newblock {Mass-radius relationships for solid exoplanets}.
\newblock \apj~669, 1279--1297.

\bibitem[{Shabalin}(2014){Shabalin}]{Shabalin2014}
{Shabalin}, I.~L., 2014.
\newblock {\em Ultra-high temperature materials I: carbon (graphene/graphite)
  and refractory metals}.
\newblock Springer.

\bibitem[{Smythe} et~al.(2017){Smythe}, {Wood}, and {Kiseeva}]{Smythe2017}
{Smythe}, D.~J., {Wood}, B.~J., {Kiseeva}, E.~S., 2017.
\newblock {The S content of silicate melts at sulfide saturation: New
  experiments and a model incorporating the effects of sulfide composition}.
\newblock American Mineralogist~102, 795--803.

\bibitem[{Southworth} et~al.(2017){Southworth}, {Mancini}, {Madhusudhan},
  {Molli{\`e}re}, {Ciceri}, and {Henning}]{Southworth2017}
{Southworth}, J., {Mancini}, L., {Madhusudhan}, N., {Molli{\`e}re}, P.,
  {Ciceri}, S., {Henning}, T., 2017.
\newblock {Detection of the atmosphere of the 1.6 M $_{⊕}$ exoplanet GJ 1132
  b}.
\newblock \aj~153, 191.

\bibitem[{Stassun} et~al.(2017){Stassun}, {Collins}, and {Gaudi}]{Stassun2017}
{Stassun}, K.~G., {Collins}, K.~A., {Gaudi}, B.~S., 2017.
\newblock {Accurate empirical radii and masses of planets and their host stars
  with Gaia parallaxes}.
\newblock \aj~153, 136.

\bibitem[{Steenstra} et~al.(2016){Steenstra}, {Knibbe}, {Rai}, and {van
  Westrenen}]{Steenstra2016}
{Steenstra}, E.~S., {Knibbe}, J.~S., {Rai}, N., {van Westrenen}, W., 2016.
\newblock {Constraints on core formation in Vesta from metal-silicate
  partitioning of siderophile elements}.
\newblock \gca~177, 48--61.

\bibitem[{Stewart} et~al.(2007){Stewart}, {Schmidt}, {van Westrenen}, and
  {Liebske}]{Stewart2007}
{Stewart}, A.~J., {Schmidt}, M.~W., {van Westrenen}, W., {Liebske}, C., 2007.
\newblock {Mars: A new core-crystallization regime}.
\newblock Science~316, 1323.

\bibitem[{Stixrude} and {Lithgow-Bertelloni}(2005){Stixrude} and
  {Lithgow-Bertelloni}]{Stixrude2005}
{Stixrude}, L., {Lithgow-Bertelloni}, C., 2005.
\newblock {Thermodynamics of mantle minerals - I. Physical properties}.
\newblock Geophysical Journal International~162, 610--632.

\bibitem[{Takahashi} et~al.(2013){Takahashi}, {Ohtani}, {Terasaki}, {Ito},
  {Shibazaki}, {Ishii}, {Funakoshi}, and {Higo}]{Takahashi2013}
{Takahashi}, S., et~al., 2013.
\newblock {Phase relations in the carbon-saturated C-Mg-Fe-Si-O system and C
  and Si solubility in liquid Fe at high pressure and temperature: implications
  for planetary interiors}.
\newblock Physics and Chemistry of Minerals~40, 647--657.

\bibitem[{Thiabaud} et~al.(2015){Thiabaud}, {Marboeuf}, {Alibert}, {Leya}, and
  {Mezger}]{Thiabaud2015}
{Thiabaud}, A., {Marboeuf}, U., {Alibert}, Y., {Leya}, I., {Mezger}, K., 2015.
\newblock {Elemental ratios in stars vs planets}.
\newblock \aap~580, A30.

\bibitem[{Toplis}(2005){Toplis}]{Toplis2005}
{Toplis}, M.~J., 2005.
\newblock {The thermodynamics of iron and magnesium partitioning between
  olivine and liquid: Criteria for assessing and predicting equilibrium in
  natural and experimental systems}.
\newblock Contributions to Mineralogy and Petrology~149, 22--39.

\bibitem[{Tsuno} and {Dasgupta}(2015){Tsuno} and {Dasgupta}]{Tsuno2015}
{Tsuno}, K., {Dasgupta}, R., 2015.
\newblock {Fe-Ni-Cu-C-S phase relations at high pressures and temperatures -
  The role of sulfur in carbon storage and diamond stability at mid- to
  deep-upper mantle}.
\newblock Earth and Planetary Science Letters~412, 132--142.

\bibitem[{Tsuno} et~al.(2007){Tsuno}, {Ohtani}, and {Terasaki}]{Tsuno2007}
{Tsuno}, K., {Ohtani}, E., {Terasaki}, H., 2007.
\newblock {Immiscible two-liquid regions in the Fe O S system at high pressure:
  Implications for planetary cores}.
\newblock Physics of the Earth and Planetary Interiors~160, 75--85.

\bibitem[{Unterborn} et~al.(2016){Unterborn}, {Dismukes}, and
  {Panero}]{Unterborn2016}
{Unterborn}, C.~T., {Dismukes}, E.~E., {Panero}, W.~R., 2016.
\newblock {Scaling the Earth: A sensitivity analysis of terrestrial
  exoplanetary interior models}.
\newblock \apj~819, 32.

\bibitem[{Unterborn} et~al.(2014){Unterborn}, {Kabbes}, {Pigott}, {Reaman}, and
  {Panero}]{Unterborn2014}
{Unterborn}, C.~T., {Kabbes}, J.~E., {Pigott}, J.~S., {Reaman}, D.~M.,
  {Panero}, W.~R., 2014.
\newblock {The role of carbon in extrasolar planetary geodynamics and
  habitability}.
\newblock \apj~793, 124.

\bibitem[{Valencia} et~al.(2006){Valencia}, {O'Connell}, and
  {Sasselov}]{Valencia2006}
{Valencia}, D., {O'Connell}, R.~J., {Sasselov}, D., 2006.
\newblock {Internal structure of massive terrestrial planets}.
\newblock \icarus~181, 545--554.

\bibitem[{Valencia} et~al.(2009){Valencia}, {O'Connell}, and
  {Sasselov}]{Valencia2009}
{Valencia}, D., {O'Connell}, R.~J., {Sasselov}, D.~D., 2009.
\newblock {The role of high-pressure experiments on determining super-Earth
  properties}.
\newblock \apss~322, 135--139.

\bibitem[{Valencia} et~al.(2007a){Valencia}, {Sasselov}, and
  {O'Connell}]{Valencia2007a}
{Valencia}, D., {Sasselov}, D.~D., {O'Connell}, R.~J., 2007a.
\newblock {Radius and structure models of the first super-Earth planet}.
\newblock \apj~656, 545--551.

\bibitem[{van Kan Parker} et~al.(2011){van Kan Parker}, {Mason}, and {van
  Westrenen}]{VanKanParker2011}
{van Kan Parker}, M., {Mason}, P.~R.~D., {van Westrenen}, W., 2011.
\newblock {Trace element partitioning between ilmenite, armalcolite and
  anhydrous silicate melt: Implications for the formation of lunar high-Ti mare
  basalts}.
\newblock \gca~75, 4179--4193.

\bibitem[{Wagner} et~al.(2011){Wagner}, {Sohl}, {Hussmann}, {Grott}, and
  {Rauer}]{Wagner2011}
{Wagner}, F.~W., {Sohl}, F., {Hussmann}, H., {Grott}, M., {Rauer}, H., 2011.
\newblock {Interior structure models of solid exoplanets using material laws in
  the infinite pressure limit}.
\newblock \icarus~214, 366--376.

\bibitem[{Wang} et~al.(1991){Wang}, {Hirama}, {Nagasaka}, and
  {Ban-Ya}]{Wang1991}
{Wang}, C., {Hirama}, J., {Nagasaka}, T., {Ban-Ya}, S., 1991.
\newblock Phase equilibria of liquid fe-s-c ternary system.
\newblock ISIJ International~31 (11), 1292--1299.

\bibitem[{Watson} et~al.(2002){Watson}, {Wark}, {Price}, and {Van
  Orman}]{Watson2002}
{Watson}, E.~B., {Wark}, D.~A., {Price}, J.~D., {Van Orman}, J.~A., 2002.
\newblock {Mapping the thermal structure of solid-media pressure assemblies}.
\newblock Contributions to Mineralogy and Petrology~142, 640--652.

\bibitem[{Whitehouse} et~al.(2018){Whitehouse}, {Farihi}, {Green}, {Wilson},
  and {Subasavage}]{Whitehouse2018}
{Whitehouse}, L.~J., {Farihi}, J., {Green}, P.~J., {Wilson}, T.~G.,
  {Subasavage}, J.~P., 2018.
\newblock {Dwarf carbon stars are likely metal-poor binaries and unlikely hosts
  to carbon planets}.
\newblock \mnras~479, 3873--3878.

\end{thebibliography}

\end{document}